\documentclass[preprint]{elsarticle}

\usepackage{lineno,hyperref}

\usepackage{subfigure}
\usepackage{citesort}
\usepackage[fleqn]{amsmath}
\usepackage{ascmac}
\usepackage{psfrag}
\usepackage{amsmath}
\usepackage{amsbsy}
\usepackage{amssymb}
\usepackage{latexsym}
\usepackage{color}
\usepackage{cases}

\newtheorem{proposition}{Proposition}

\newtheorem{theorem}{Theorem}
\newtheorem{lemma}{Lemma}

\newtheorem{assump}{Assumption}

\newtheorem{remark}{Remark}

\newcommand{\migip}{\hspace*{\fill} $\Box $}

\newcommand{\signal}[1]{{\boldsymbol{#1}}}
\newcommand{\Natural}{{\mathbb N}}
\newcommand{\norm}[1]{\left\|#1\right\|}
\newcommand{\abs}[1]{\left|#1\right|}

\newcommand{\real}{{\mathbb R}}
\newcommand{\comp}{{\mathbb C}}
\newcommand{\innerprod}[2]{\left\langle{#1},{#2}\right\rangle}

\newcommand{\refeq}[1]{(\ref{#1})}

\newcommand{\argmin}{\operatornamewithlimits{argmin}}

\graphicspath{{./images/}}

\journal{Signal Processing}

\begin{document}

\begin{frontmatter}

\title{Relaxed Zero-Forcing Beamformer under Temporally-Correlated Interference}
\tnotetext[mytitlenote]{This work was partially supported by JSPS Grants-in-Aid 18H01446.
A preliminary version of this paper was presented at 44th IEEE International Conference on
Acoustics, Speech, and Signal Processing, Brighton, UK, May 2019.
The first and second authors have an equal contribution.}

\author{Takehiro Kono$^a$,  Masahiro Yukawa$^{a*}$}
\affiliation{organization={Department of 
Electronics and Electrical Engineering, Keio University},
            addressline={Hiyoshi 3-14-1, Kohoku-ku}, 
            city={Yokohama},
            postcode={223-8522}, 
            country={JAPAN}}
\cortext[mycorrespondingauthor]{Corresponding author: Masahiro Yukawa (yukawa@elec.keio.ac.jp)}

\author{Tomasz Piotrowski$^b$}
\affiliation{organization={Interdisciplinary Center for Modern Technologies
and the Faculty of Physics, Astronomy and Informatics, Nicolaus Copernicus
University},
            city={Torun},
            country={POLAND}}




\begin{abstract}
The relaxed zero-forcing (RZF) beamformer is a
quadratically-and-linearly constrained minimum variance beamformer.  The
central question addressed in this paper is whether RZF performs better
than the widely-used minimum variance distortionless response and
zero-forcing beamformers under temporally-correlated interference.
First, RZF is rederived by imposing an ellipsoidal constraint that
bounds the amount of interference leakage for mitigating the intrinsic
gap between the output variance and the mean squared error (MSE) which
stems from the temporal correlations.  Second, an analysis of RZF is
presented for the single-interference case, showing how the MSE is
affected by the spatio-temporal correlations between the desired and
interfering sources as well as by the signal and noise powers.  Third,
numerical studies are presented for the multiple-interference case,
showing the remarkable advantages of RZF in its basic performance as
well as in its application to brain activity reconstruction from EEG
data.
 The analytical and experimental results clarify that the
RZF beamformer gives near-optimal performance in some situations.

\end{abstract}


\end{frontmatter}


\section{Introduction}\label{sec:intro}

Beamforming is an attractive technology to filter out those interfering
(or jamming) signals
which are spatially apart from the signal of interest.
It has a long history (see e.g., \cite{trees02_book}) and has been applied
to a wide range of applications including radar, sonar, 
wireless communications, acoustics, electroencephalography (EEG), among many others.
For the upcoming fifth generation (5G) wireless network and Internet of
things (IoT) technologies, for instance, beamforming is considered as a
key component to revolutionize connectivity
as well as massive MIMO (multiple input multiple output) and
millimeter wave \cite{sun14,taiwo19,bana19}.
The minimum-variance distortionless response (MVDR) beamformer minimizes
the variance of the beamformer output under a linear constraint to
preserve the desired signal undistorted \cite{van1997localization}.
The MVDR beamformer is known to achieve
the optimal performance among all linear beamformers
in the absence of correlations among sources.
In many applications, however, the sources have {\em temporal correlation},
which makes MVDR no longer optimal
\cite{hui2010identifying}.
Throughout the paper, unless otherwise stated explicitly, the term
``correlation'' means a temporal correlation between the desired and
interfering sources.
The zero-forcing (ZF) beamformer (aka.~the nulling beamformer)
\cite{dalal2006modified,hui2010identifying,hui2006linearly,piotrowski19}
minimizes the output variance under multiple linear constraints
to preserve the desired signal undistorted and, at the same time,
to nullify the interfering signals.
Although it could attain better performance than MVDR in the correlated-source
settings with high signal to noise ratio (SNR),
its performance becomes even worse than MVDR in low SNR cases
as it amplifies the noise \cite{hui2006linearly}.
The MVDR and ZF beamformers are particular examples of the linearly constrained
minimum variance (LCMV) beamformer \cite{Frost72:ProcIEEE,van1997localization},
and can be implemented by efficient adaptive algorithms
such as the constrained normalized least mean square (CNLMS) algorithm
\cite{apolinario1998constrained}.


The studies of 
the temporal correlation between the desired and interfering signals
date back
at least to 1970's \cite{owsley74} in the communication area.
The major causes of the correlation are multipath propagation
which may also happen in other applications such as acoustics)
and/or ``smart'' jammers which aim to transmit antiphase replicas of 
the desired signal for jamming the transmission.
In the worst case, 
the desired and interfering signals would be coherent; i.e., one source
signal is an amplitude-scaled phase-shifted version of another one,
or in other words the correlation coefficient has a unit amplitude.
To alleviate this problem, {\em spatial smoothing} and its improved
versions have been proposed, and the impacts of the correlation
coefficient on the performance have been investigated in the literature
(see \cite[Section 6.12]{trees02_book} and the references therein).
This technique, however, has several limitations.
For instance, it is only applicable to a regular array geometry such as
uniform linear arrays, and/or
it requires twice as many sensors as signal sources.
An efficient beamformer design that can manage the correlated
interference would be beneficial in the future wireless network systems
to enhance the connectivity.

Another primary application of beamformers is in inverse problems in EEG
and magnetoencephalography (MEG), which are imaging modalities that provide
a direct measure of the brain activity by measuring the voltage 
with a set of sensors placed at various locations on the scalp of the subject.
The EEG inverse problem aims to localize and reconstruct the sources of
brain electrical activity from the EEG measurements.
It is widely known that neighboring brain activities (sources) have strong
correlations typically, and the signal correlation still
remains a challenging issue to be addressed 
in the brain signal processing area.
Focusing on EEG, this is especially true for applications requiring online processing of
beamformer output such as brain-computer interface (BCI)
\cite{Wittevrongel2016BCI,Mahmoodi2017BCI} and source-space EEG
neurofeedback, which is currently another hot topic in brain signal
processing \cite{Boe2014neurofeedback,Gepper2017neurofeedback}.
To date, the dominant beamforming technique used in these fields has been
the MVDR beamformer, see, e.g.,
\cite{Mullen2015BCI,Hartmann2014neurofeedback,vanLutterveld2017neurofeedback}.
Therefore, an introduction of a beamformer design which suppresses efficiently
interfering activity, includes as special cases the MVDR and ZF
beamformers, and is amenable to efficient adaptive implementation,
should bring significant benefits to BCI and source-space EEG
neurofeedback.



Let us now turn our attention to 
the relaxed zero forcing (RZF) beamformer
\cite{yukawa2011adaptive,yukawa2013dual}.
Despite its close link to the ZF beamformer,
RZF has solely been studied under the assumption that
the source signals are mutually {\em uncorrelated} in time domain.
In this case, MVDR  is the optimal linear beamformer,
and hence no other linear beamformer can outperform it.\footnote{
For clarification, the term ``linear'' here means that the beamformer output
is linear in terms of the received vector (the input to the beamformer).
RZF is a ``linear beamformer'' involving a ``nonlinear (quadratic)
constraint'' in its formulation.}
In fact, the key point of the previous works of RZF lies in 
its adaptive implementation.
The central theme addressed therein is an efficient use of
the side information about the interference channel 
for enhancing the convergence speed of adaptive algorithms.
The side information is accommodated in the RZF formulation
as an ellipsoidal constraint that bounds the contribution of
interference in the output
using the interference channel information.
The RZF beamformer contains MVDR and ZF as its two extremes:
MVDR does not consider the correlation explicitly, while
ZF annihilates the interfering signals completely.
The RZF beamformer belongs to the class of
linearly and quadratically constrained minimum variance (LQCMV)
beamformer
due to the presence of the quadratic constraint as well as the typical
distortionless constraint in its formulation.
The classical adaptive algorithms such as CNLMS or its projective extension using
the projection onto a nonlinear constraint set
cannot be applied to RZF since the projection onto the
ellipsoidal constraint set admits no closed-form expression.
Fortunately, however, it can be implemented efficiently
by the multi-domain adaptive learning method
\cite{ysy_ieice10_multi}, which implements the ellipsoidal constraint
via the projection onto a closed ball in a transform domain.
The multi-domain method applied to RZF is referred to as 
the dual-domain adaptive algorithm (DDAA)
\cite{ysy_ieice10_multi,yukawa2013dual}, in which
the side information plays a role of guiding
the update direction of adaptive beamformer towards
the optimal one and thereby yielding
remarkably fast convergence.
As mentioned already, RZF has not yet been studied
under the correlated interference which permits
some other linear beamformer to be superior to MVDR.
For clarification, the superiority is discussed here in terms of
the mean squared error (MSE) performance {\em in the batch setting},
{\em not} in terms of convergence speed of adaptive algorithms.
Our central research question is whether
the RZF beamformer can be superior to MVDR and ZF, and,
if the answer is positive, under what conditions it may happen.

The major contributions of the present paper are summarized below.
\begin{enumerate}
 \item A rederivation of the RZF beamformer is presented
based on a decomposed form of the MSE function
in the presence of temporally correlated interference.
 \item A theoretical analysis of the RZF beamformer is presented
for the tractable case of a single interfering signal that is
       correlated with the desired signal.
Here, the analysis relies on the isomorphism
       between $\comp^2$ and
the two dimensional subspace spanned by the channel vectors of the
       desired and interfering sources
under some mapping preserving the norms and inner products.

 \item The performance of the RZF beamformer is studied through
extensive simulations for the general case of multiple
correlated interfering signals.
 \item Consequently, 
the RZF beamformer is shown to  outperform the
MVDR and ZF beamformers, in the batch setting,
in the presence of correlated interference also in an EEG setup, where the beamformers are used for online reconstruction of brain bioelectrical activity from realistically simulated EEG measurements.
It should be emphasized that this is a remarkable difference
from the previous studies
       \cite{yukawa2011adaptive,yukawa2013dual}
for uncorrelated interference, which only allows RZF to outperform
MVDR in terms of the convergence speed of adaptive algorithms.

\end{enumerate}

The MSE function can actually be decomposed into the following three terms:
(i) the output variance, 
(ii) the desired-signal-power dependent term,
and (iii) the interaction with interference
(which will be referred to as the interference term).
The interference term here depends on the correlation
between the desired and interfering signals
(as well as the beamformer outputs of the interference channel vectors).
No prior knowledge is assumed available about the correlation,
and the primitive idea is mitigating (to some reasonable extent),
but {\it not completely annihilating}, the interference term.
Although the interference term cannot be suppressed directly, 
it is mitigated indirectly by a certain quadratic constraint
that bounds the total power of the beamformer outputs for the
interference channel vectors.
The resultant beamformer coincides with the RZF beamformer,
although its motivation as well as the underlying assumption
is totally different from the original work
\cite{yukawa2011adaptive,yukawa2013dual}.
The MSE of RZF for the single-interference case is analyzed,
and its explicit formula is derived
in terms of the spatio-temporal correlations between the desired
and interference sources as well as the variances of noise and interference.
The analysis reveals that RZF achieves the minimum MSE (MMSE)
under certain conditions, and it also clarifies under what conditions RZF
outperforms MVDR and ZF, respectively.
Numerical examples show that the RZF beamformer achieves significant gains
over the whole range of SNR and that its performance is fairly close to
the theoretical bound for high SNRs, and it significantly outperforms the MVDR and ZF
beamformers.
To examine the possibility of exploiting the temporal
correlations to enhance the performance further,
we also investigate the approximate minimum MSE (A-MMSE) beamformer
relying on the availability of some erroneous estimates
of the correlations between the desired and interfering sources
as well as the availability of an estimate of the signal power.
It turns out that A-MMSE is sensitive to
the errors in those estimates,
while RZF is fairly insensitive to the choice of the
power-bounding parameter.

We emphasize that the present study is the first work that investigates the RZF beamformer
in the correlated-interference scenario, 
demonstrating its efficacy in managing such interference.
Compared to the preliminary version \cite{kono_icassp19},
the new features of this paper include
(i) treatment of complex signals,
(ii) comprehensive treatment of beamforming problems (rather than
focusing on the specific EEG application),
(iii) theoretical studies of RZF (Section \ref{sec:analysis}), and 
(iv) extensive simulation studies including an EEG application,
together with considerations of the A-MMSE beamformer
(Section \ref{sec:exp}).

\section{System Model and Assumptions}
\label{sec:prelim}
Throughout, $\mathbb{R}$, $\comp$, and $\mathbb{N}$ denote the sets
of real numbers, complex numbers, and nonnegative integers,
respectively.
Given a matrix $\signal{A}$, $\signal{A}^{\sf T}$ and
$\signal{A}^{\sf{H}}$ denote
its transpose and  Hermitian transpose, respectively, and 
$\sigma_{{\rm max}}(\signal A)$ denotes its largest singular value. 
The identity matrix is denoted by
$\signal{I}$, and the zero vector by $\signal{0}$.
Given vectors $\signal{x},\signal{y} \in
\comp^{m}$ of arbitrary dimension $m \in \mathbb{N}^{*} :=
\mathbb{N}\setminus\{ 0 \}$, 
define the inner product by $\langle {\signal
x},\signal{y} \rangle := \signal{x}^{\sf H}\signal{y}$, and the induced
norm by 
$\|\signal{x}\| := \langle
\signal{x},\signal{x} \rangle^{\frac{1}{2}}$. 
The expectation is denoted by 
$E(\cdot)$.

We consider the situation when $J+1$ spatially-separated sources transmit
signals towards a uniform linear array\footnote{
Different types of antenna arrays have also been studied such as 
nonuniform linear arrays, 
nested array,
coprime arrays, or 
the maximum inter-element spacing constraint (MISC) arrays (see, e.g.,
\cite{hoctor90, piya10, liu16, zhou18, zheng19, li20}).
An extension to those different arrays is beyond the
scope of the current work.}
with $N$ antenna sensors.
The received signal at sensors at time instant $k\in\mathbb{N}$ is
modeled as
\begin{eqnarray}
\signal{y}(k)=\sum_{j=0}^{J} s_j(k) \signal{h}_j +
 \signal{n}(k) \in \comp^{N}.
\label{eq:eeg_model}
\end{eqnarray}
Here, $\signal{h}_j:=[h_{j,1},h_{j,2},\cdots,h_{j,N}]^{\sf T} \in \comp^{N}$ is
the channel vector of the $j$th source such that $\norm{\signal{h}_j}=1$,
$s_j(k)\in \comp$ the $j$th transmit signal, and
$\comp^{N}\ni\signal{n}(k):=[n_1(k),n_2(k),\cdots,n_N(k)]^{\sf T} \sim \mathcal{CN}
(\signal{0},\sigma_n^2\signal{I}) $ the noise vector.
Without loss of generality, $s_0(k)$ is supposed to be the desired
signal, and all the others are interference.
The goal of beamforming is to reconstruct the desired signal $s_0(k)$
from the measurements $\signal{y}(k)$ with the channel information.
The assumptions that will be used in the current study are listed below.

\begin{assump}\label{assmuption:signal}
{}~
 \begin{enumerate}
  \item All the signals $s_j(k)$ and noise are zero-mean
weakly-stationary stochastic processes.
  \item The interfering signals $s_j(k)$, $j=1,2,\cdots,J$,
 are correlated with the desired signal $s_0(k)$, i.e.,
\begin{equation}
c_j:=E[s_0^*(k)s_j(k)]\neq 0, ~~~j=1,2,\cdots,J.
\end{equation}
  \item The noise is uncorrelated with
the signals, i.e.,  $E[s_j^*(k)\signal{n}(k)] = \signal{0}$ for $j=0,1,\cdots,J$.
  \item All the channel vectors $\signal{h}_j$ are
	available.\footnote{This is a realistic assumption 
in uplink communication in wireless
networks and also in EEG applications.}
\item The correlations $c_j$ and the signal power $\sigma_0^2:=E(\abs{s_0(k)}^2)$
are completely unknown.
 \end{enumerate}
\end{assump}

We note that the model in \refeq{eq:eeg_model} actually gives 
the forward model in the EEG/MEG inverse problem, 
which aims to reconstruct the activity of the desired
source $s_0(k)$ from the EEG/MEG measurements $\signal{y}(k)$.
In this interpretation of the model in \refeq{eq:eeg_model}, for each source $j$, the channel vector 
$\signal{h}_j:=\signal{h}(\signal{r}_j,\signal{u}_j)\in\real^{N}$
corresponds to the so-called leadfield vector depending on
the source position  $\signal{r}_{j}$ and 
the orientation unit vector $\signal{u}_{j}$,
$s_j(k)\in \real$ is the electric/magnetic dipole moment
at time instant $k$, and
the noise vector $\signal{n}(k) \in \real^{N}$ represents
the background brain activity along with the noise
recorded at the sensor array. We will use this fact in Section \ref{subsec:eeg}, where the performance of beamformers under consideration will be evaluated in this context.

\section{Relaxed Zero-forcing Beamformer Rederived to Manage Temporally-correlated Interference}
\label{sec:rzf}

Under Assumption \ref{assmuption:signal}, 
the MSE function can be written as
\begin{align}
\hspace*{-2em}J_{{\rm MSE}}(\signal{w}) :=&~
E\left[\abs{\signal{w}^{\sf H}\signal{y}(k) - s_0(k)}^2\right]\nonumber\\
=&~ \signal{w}^{\sf H} \signal{R} \signal{w} + 
\sigma_0^2(1-\signal{w}^{\sf H}\signal{h}_0-\signal{h}_0^{\sf H}\signal{w})
  - \sum_{j=1}^{J} (c_j \signal{w}^{\sf H}\signal{h}_j + c_j^*
 \signal{h}_j^{\sf H}\signal{w}),
~\signal{w}\in \comp^N,
\label{eq:mse_cost}
\end{align}
where 
$\signal{R}:=E[\signal{y}(k)\signal{y}(k)^{\sf H}]$.
Under the distortionless constraint
\begin{equation}
 \signal{w}\in C := \{\signal{w}\in \comp^{N} \mid \signal{w}^{\sf
  H}\signal{h}_0 = 1\},
\label{eq:def_C}
\end{equation}
\eqref{eq:mse_cost} reduces to
\begin{equation}
J_{{\rm MSE}}(\signal{w}) 
= \underbrace{\signal{w}^{\sf H} \signal{R} \signal{w}}_{\mbox{output variance}} - 
\underbrace{\sigma_0^2}_{\mbox{signal power}} 
 - \underbrace{\sum_{j=1}^{J} (c_j \signal{w}^{\sf H}\signal{h}_j +
 c_j^* \signal{h}_j^{\sf H}\signal{w}).}_{\mbox{interaction with
 interference}}\label{eq:mse_cost_dr}
\end{equation}
In the absence of correlated sources,
the third term of \refeq{eq:mse_cost_dr} disappears, and hence
MSE (under the distortionless constraint) can be minimized by minimizing
the output variance.
This is the reason why the MVDR beamformer, the minimizer of
\eqref{eq:mse_cost_dr} under \eqref{eq:def_C}, is optimal.
In the presence of correlated sources, on the other hand,
the third term makes significant differences and MVDR is no longer optimal.
In this case, one may consider to annihilate the third term by imposing
the nulling constraints
$\signal{w}^{\sf H}\signal{h}_j=0$ for all $j=1,2,\cdots,J$.
This is the so-called ZF beamformer.
Unfortunately, the ZF beamformer is known to amplify the noise, and
its performance degrades seriously under noisy environments.
To circumvent the noise amplification problem while exploiting the
channel information of the interfering signals, 
we impose a quadratic constraint, 
in addition to the linear constraint \eqref{eq:def_C},
that bounds the amount of interference leakage
by a certain threshold $\varepsilon \geq 0$,
rather than annihilating those signals completely.
The RZF beamformer is formally given as follows:
\begin{subequations}
\label{eq:rzf}
    \begin{align}
\hspace*{-2em}&{\rm minimize~~~} \signal{w}^{\sf H} \signal{R} \signal{w}
 \label{eq:rzf1}\\
\hspace*{-2em}&{\rm ~subject ~ to~}\left\{  
\begin{array}{l}
\signal{w}^{\sf H}\signal{h}_0 = 1 ~~~(\Leftrightarrow \signal{w}\in
 C),\\
\|\signal{H}_{\rm I}^{\sf H}\signal{w}\|^{2}\leq \varepsilon ~~~
(\Leftrightarrow \signal{H}_{\rm I}^{\sf H} \signal{w}\in  B_{\varepsilon}
:=\{\signal{s}\in \comp^{J} \mid \|\signal{s}\|^{{\rm 2}}\leq \varepsilon\}),
\end{array}\right.
\label{eq:rzf2}
    \end{align}
\end{subequations}
where $\signal{H}_{\rm I} := [\signal{h}_1~\signal{h}_2\cdots
\signal{h}_J]$ is 
the interference channel matrix.
Let
$\signal{R}_{\varepsilon}:= \signal{R} + \lambda_{\varepsilon}\signal{H}_{\rm
I}
\signal{H}^{\sf H}_{\rm I}$ with
the Lagrange multiplier
$\lambda_{\varepsilon}\geq 0$ satisfying
$\norm{\signal{H}^{\sf H}_{\rm I}\signal{w}_{\rm RZF}}^2=
\norm{ \dfrac{\signal{H}_{\rm I}^{\sf H} (\signal{R} + \lambda_{\varepsilon}\signal{H}_{\rm I}\signal{H}^{\sf H}_{\rm I})^{{\rm -1}}{\signal h}_{{\rm 0}}}
{\signal{h}_{{\rm 0}}^{\sf H} 
(\signal{R} + \lambda_{\varepsilon}\signal{H}_{\rm I}\signal{H}^{\sf H}_{\rm I})^{{\rm -1}}
\signal{h}_{{\rm 0}}}}^2=\varepsilon$,
$\signal{H}:=[\signal{h}_0~\signal{H}_{\rm I}]\in\comp^{N\times (J+1)}$, and
$\signal{c}:=[1,\signal{0}^{\sf T}]^{\sf T}\in\{0,1\}^{J+1}$.
The solution to the problem in \refeq{eq:rzf}
is then given by \cite{yukawa2013dual}
\begin{equation}
 \signal{w}_{\rm RZF}=\begin{cases}
\dfrac{\signal{R}_{\varepsilon}^{{\rm -1}}{\signal h}_{{\rm 0}}}
{\signal{h}_{{\rm 0}}^{\sf H}\signal{R}_{\varepsilon}^{{\rm
 -1}}\signal{h}_{{\rm 0}}},	 & \mbox{ if } \varepsilon>0, \\
\signal{w}_{{\rm ZF}}:=\signal{R}^{-1} \signal{H}\left(\signal{H}^{\sf
			 H}\signal{R}^{-1}\signal{H}\right)^{-1}\signal{c},
			 & \mbox{ if } \varepsilon=0.
			\end{cases}
\label{eq:wrzf_analytic_solution}
\end{equation}
For the particular choice of 
$\lambda_{\varepsilon}:=0$,
RZF reduces to the MVDR beamformer $\signal{w}_{\rm MVDR}:= 
\dfrac{ \signal{R}^{{\rm -1}}{\signal h}_{{\rm 0}}}
{\signal{h}_{{\rm 0}}^{\sf H} \signal{R}^{\rm -1}\signal{h}_{{\rm 0}}}$,
of which the corresponding $\varepsilon$ is
 $\varepsilon_{\rm MVDR}:= 
\norm{ \signal{H}_{\rm I}^{\sf H} \signal{w}_{\rm MVDR}}^2$.
In contrast, as $\lambda_{\varepsilon}\rightarrow + \infty$, the
corresponding $\varepsilon$ vanishes.
In fact, $\lambda_{\varepsilon}\in[0,+\infty)$ is a strictly monotone decreasing
function of $\varepsilon\in (0,\varepsilon_{\rm MVDR}]$, and thus the correspondence
is one to one.

The RZF beamformer attains significantly better performance than MVDR
and ZF and could achieve (near) optimal performance in some cases,
as shown in Sections \ref{sec:analysis} and \ref{sec:exp}.
For self-containedness, an adaptive implementation of RZF
by the dual-domain adaptive algorithm \cite{ysy_ieice10_multi} is given
in the appendix.

\begin{remark}[On Unbiasedness]
\label{remark:unbiasedness}


Let $\mu_j\in\comp$ denote the mean value of $s_j(k)$;
 i.e., $\mu_j:=E[s_j(k)]$, $j=0,1,\cdots,J$.
Although $\mu_j:=0$, $j=0,1,\cdots,J$, is assumed in Assumption
 \ref{assmuption:signal}, we also consider the general case here
to clarify the statistical assumptions required for unbiasedness.
We discuss the unbiasedness of a linear estimator
 $\widehat{s}_0(k):=\signal{w}^{\sf H} \signal{y}(k)$ of 
a random variable $s_0(k)$
with a beamforming vector $\signal{w}$ subject to 
the distortionless constraint $\signal{w}^{\sf H} \signal{h}_0=1$.
This includes the case of the RZF beamformer in \eqref{eq:rzf}.
Inserting
\eqref{eq:eeg_model} into 
$\widehat{s}_0(k)=\signal{w}^{\sf H} \signal{y}(k)$ yields
\begin{equation} \label{reveal}
\widehat{s}_0(k)=\signal{w}^{\sf H} \signal{h}_0 s_0(k)+\sum_{j=1}^J
 \signal{w}^{\sf H} \signal{h}_j s_j(k)+ \signal{w}^{\sf H} \signal{n}(k).
\end{equation}
Since $s_0(k)$ is modeled as a random variable, the unbiasedness
condition of the estimator $\widehat{s}_0(k)$ is given by
\cite{kailath00}:
\begin{equation*}
E[\widehat{s}_0(k)] = \mu_0.
\end{equation*}
Due to the zero-mean assumption
$E[\signal{n}(k)]=\signal{0}$ for noise
as well as the distortionless constraint $\signal{w}^{\sf H} \signal{h}_0=1$,
it follows that
\begin{equation}
E[\widehat{s}_0(k)] = \mu_0+\sum_{j=1}^J \signal{w}^{\sf H} \signal{h}_j\mu_j.
\end{equation}
This means that $\widehat{s}_0(k)$ is an unbiased estimator of $s_0(k)$ 
if $\sum_{j=1}^J\signal{w}^{\sf H} \signal{h}_j\mu_j=0$.
We thus obtain the following observations.
\begin{enumerate}
 \item Suppose that all interfering signals have zero mean, i.e., if
       $\mu_j(:=E[s_j(k)])$
$=0$ for $j=1,2,\dots,n$.
In this case, \emph{any} estimator of the form 
$\widehat{s}_0(k):=\signal{w}^{\sf H} \signal{y}(k)$
is unbiased under the distortionless constraint.
This applies to the present study, since the conditions are clearly satisfied
under Assumption \ref{assmuption:signal}.
Note here that the only statistical assumptions required to draw
this conclusion is the zero-mean assumptions of the interfering signals
       and the noise vector.

 \item Suppose that some interfering signal(s) has (have) non-zero mean.
In this case, 
the unbiasedness condition $\sum_{j=1}^J\signal{w}^{\sf H}
       \signal{h}_j\mu_j=0$
can be achieved, e.g.,
by using the zero-forcing constraints $\signal{w}^{\sf H} \signal{h}_j=0$ for
$j=1,\dots,n$, or only for those interfering signals with non-zero means (if known).

\end{enumerate}
\end{remark}

\begin{remark}[Tips for $\epsilon$-Parameter Design]
\label{remark:epsilon_design}

Below are some tips on how to choose the  $\epsilon$ parameter.
\begin{itemize}
 \item In the high SNR regime, $\epsilon$ should be set to a small
       value, since the impact of noise on MSE is small
(see \eqref{eq:jmse_distortionless} in Lemma \ref{lemma:mse}).

 \item When one (or more) of the interfering signals, say the first one,
is spatially correlated to the desired signal, i.e.,
$\abs{\innerprod{\signal{h}_0}{\signal{h}_1}}\approx 1$, 
$\epsilon$ should be set to a large value.
The reason is the following: if $\epsilon$ is small in such a situation,
the beamforming vector $\signal{w}$ needs to be nearly orthogonal to the 
interference channel vectors including $\signal{h}_1$,
i.e., $\innerprod{\signal{w}}{\signal{h}_1}\approx0$,
but at the same time it also needs to satisfy $\innerprod{\signal{w}}{\signal{h}_0}=1$,
and those two requirements under $\abs{\innerprod{\signal{h}_0}{\signal{h}_1}}\approx 1$
make the norm $\norm{\signal{w}}$
       unacceptably large
and thus cause serious noise amplification.
 
\end{itemize}
Adaptation of $\epsilon$ is left as an interesting open issue.

\end{remark}

\section{Analytical Study of RZF Beamformer for Single Interference Case}
\label{sec:analysis}

After some general discussions about the MSEs of beamformers,
we present the MSE analysis of RZF for the single-interference case.
The proofs are finally given together with some remarks.

\subsection{General Discussions}

It is straightforward to verify the following lemma.
\begin{lemma}
\label{lemma:mse}
Let $c_{l,j}:=E(s_l^*(k)s_j(k))$, $l,j=1,2,\cdots J$.
Then, the MSE defined in \eqref{eq:mse_cost} 
can be rewritten as
\begin{align}
\hspace*{-0em} J_{\rm MSE}(\signal{w})= &~
\sigma_0^2 \left(1+\abs{\signal{w}^{\sf H}\signal{h}_0}^2 -
\signal{w}^{\sf H}\signal{h}_0 - \signal{h}_0^{\sf H}\signal{w} \right)
+ \sigma_n^2 \norm{\signal{w}}^2
\nonumber\\
\hspace*{-0em}&+\sum_{j=1}^{J} c_j (\signal{h}_0^{\sf H}\signal{w} -1)
\signal{w}^{\sf H}\signal{h}_j
+\sum_{j=1}^{J} c_j^* (\signal{w}^{\sf H}\signal{h}_0 -1)
\signal{h}_j^{\sf H}\signal{w}
\nonumber\\
\hspace*{-0em}&+\sum_{l=1}^{J}\sum_{j=1}^{J}c_{l,j} \signal{h}_l^{\sf H}
 \signal{w} \signal{w}^{\sf H} \signal{h}_j, ~~~\signal{w}\in \comp^N.
\end{align}
Under the distortionless constraint \eqref{eq:def_C},
MSE reduces to
\begin{equation}
 J_{\rm MSE}(\signal{w})= 
\sigma_n^2 \norm{\signal{w}}^2+\sum_{l=1}^{J}\sum_{j=1}^{J}c_{l,j} 
\signal{h}_l^{\sf H} \signal{w} \signal{w}^{\sf H}
 \signal{h}_j,~\signal{w}\in C.
\label{eq:jmse_distortionless}
\end{equation}
The minimizer of \eqref{eq:jmse_distortionless} is given by
\begin{equation}
 \signal{w}_{\rm
  MMSE\mathchar`-DR}:=\frac{\widetilde{\signal{R}}^{-1}\signal{h}_0}
{\signal{h}_0^{\sf H}\widetilde{\signal{R}}^{-1}\signal{h}_0},
\label{eq:mmsedr}
\end{equation}
which will be referred to
shortly as 
MMSE-DR (MMSE under distortionless response constraint),
where $\widetilde{\signal{R}}:=
\sigma_n^2 \signal{I} +
\sum_{l=1}^{J}\sum_{j=1}^{J}c_{l,j} 
 \signal{h}_j\signal{h}_l^{\sf H}$.
\end{lemma}

In the analysis of the RZF beamformer,
the following observation is useful:
\begin{equation}
 \signal{w}_{\rm RZF}\in {\rm span}\{\signal{h}_j\}_{j=0}^J:=\left\{
\sum_{j=0}^{J}\zeta_j
  \signal{h}_j\mid \zeta_j\in \comp\right\},
\label{eq:wrzf_property}
\end{equation}
which can easily be verified by 
applying the matrix inversion lemma \cite{horn_johnson85} 
twice in \eqref{eq:wrzf_analytic_solution}, 
first to $\signal{R}_{\varepsilon}^{-1}$
and then to $\signal{R}^{-1}$ which appears
in the outcome of the first application of the lemma.
In the rest of this section, we restrict our attention to 
the tractable case of $J:=1$.
In this case, \eqref{eq:wrzf_property} implies that
$\signal{w}_{\rm RZF}$ belongs to the two dimensional subspace 
${\rm span}\{\signal{h}_0,\signal{h}_1\}$,
and hence the problem reduces completely to the two dimensional case
of $N:=2$ by considering the isomorphism between
the two dimensional subspace ${\rm span}\{\signal{h}_0,\signal{h}_1\}$
of $\comp^N$ and $\comp^2$
under some mapping preserving the norms and inner products.
We shall discuss first about the real case, and 
then move to the complex case.

\subsection{Real Case for $J:=1$}

We present analyses of RZF for the real-signal case
under the presence of single interference below.

\begin{theorem}
\label{theorem:rzf_real_Nd}
Let $J:=1$, $\norm{\signal{h}_0}=\norm{\signal{h}_1}=1$,
$\innerprod{\signal{h}_0}{\signal{h}_1}=\sin \tau\in(-1,1)$,
the spatial-separation factor
$\tau\in(-\pi/2,\pi/2)$,
the powers
$\sigma_0^2>0$,
$\sigma_1^2>0$,
$\sigma_n^2>0$
of the desired signal $s_0$, the interference $s_1$, and noise, respectively,
and 
the correlation factor
$c_1(:=E[s_0^*(k)s_1(k)])
\in[-\sigma_0\sigma_1,\sigma_0\sigma_1]$.\footnote{The correlation coefficient
satisfies $\rho_1:=c_1/\sigma_0\sigma_1\in [-1,1]$ in general.}
Note here that the trivial case of $\sin\tau=\pm 1$ is excluded,
since there is no spatial separation between
$\signal{h}_0$ and $\signal{h}_1$ in this case.
The MSE of the RZF beamformer is given,
as a function of the Lagrange multiplier $\lambda_\varepsilon$,
by
\begin{equation}
 {\rm MSE}(\lambda_\varepsilon) = \frac{\delta^2(\sigma_1^2 \cos^2\tau +
  \sigma_n^2)}{g^2(\lambda_{\varepsilon})}
- \frac{2\sigma_n^2 \delta\tan\tau}{g(\lambda_{\varepsilon})}
+ \sigma_n^2(\tan^2\tau +1),
\label{eq:theorem_mse}
\end{equation}
where
\begin{align}
 \delta:=&~\sigma_n^2\tan \tau-c_1 \cos\tau, \label{eq:def_delta}\\
 g(\lambda_{\varepsilon}):=&~\cos^2\tau \lambda_\varepsilon + \sigma_1^2
  \cos^2\tau+\sigma_n^2>0.
\label{eq:def_g}
\end{align}

\begin{enumerate}
 \item Assume that $\delta=0$.
Then, the MSE reduces to
       $\sigma_n^2 (\tan^2\tau+1)$, which is constant in $\lambda_{\varepsilon}$.

 \item Assume that $\delta\neq0$.
If 
\begin{equation}
 \gamma:=\dfrac{\sigma_n^2\tan\tau}{\delta}\leq 0,
\label{eq:def_gamma}
\end{equation}
${\rm MSE}(\lambda_\varepsilon)$ is monotonically decreasing, and 
the ZF beamformer (corresponding to the limit of
$\lambda_\varepsilon\rightarrow +\infty$) is optimal in the MSE sense.
If
$\gamma> 0$ ($\Rightarrow \sin\tau\neq 0$),
${\rm MSE}(\lambda_\varepsilon)$  is minimized by
\begin{equation}
 \lambda_{\varepsilon}:=
\begin{cases}
-\dfrac{c_1(\sigma_1^2\cos^2\tau + \sigma_n^2)}{\sigma_n^2\sin\tau}>0, & \mbox{if } \gamma\in(0,1),\\
0, & \mbox{if } \gamma\geq 1.
\end{cases}
\label{eq:theorem_lambda}
\end{equation}
Note here by \refeq{eq:def_delta} and \eqref{eq:def_gamma} that
       $0<(\gamma^{-1}-1)/\cos^2 \tau = -c_1/(\sigma_n^2\sin\tau)$ for $\gamma\in(0,1)$.

\end{enumerate}

\end{theorem}

\begin{theorem}[Achieved MSE in real case]
\label{theorem:real_mse}
Assume that $\delta(:=\sigma_n^2\tan \tau-c_1 \cos\tau) = 0$.
Then, it holds that
$\mbox{MSE}_{\rm RZF}= \mbox{MSE}_{\rm ZF}=\mbox{MSE}_{\rm MVDR}$
for any $\lambda_\varepsilon\geq 0$.
Assume now that $\delta\neq 0$.
Then, with the optimal 
Lagrange multiplier
$\lambda_\varepsilon$, it holds that
\begin{equation}
  \mbox{MSE}_{\rm RZF}= 
\begin{cases}
  \mbox{MSE}_{\rm ZF},
& \mbox{if } \gamma\left(:=(\sigma_n^2\tan\tau)/\delta\right)\leq 0,\\
  \mbox{MSE}_{\rm MMSE\mathchar`-DR},
& \mbox{if } \gamma\in(0,1),\\
  \mbox{MSE}_{\rm MVDR},
& \mbox{if } \gamma\geq 1,
\end{cases}
\label{eq:mse_rzf_real}
\end{equation}
where
\begin{align}
 \mbox{MSE}_{\rm MVDR} =&~ \frac{\sigma_n^2(\sigma_1^2+\sigma_n^2)
 +\abs{c_1}^2\cos^2\tau}{\sigma_1^2\cos^2\tau+\sigma_n^2}, \label{eq:mse_mvdr_real}
\\
 \mbox{MSE}_{\rm ZF} =&~ \sigma_n^2(\tan^2\tau +1), \label{eq:mse_zf_real}
\\
\mbox{MSE}_{\rm MMSE\mathchar`-DR}
 =&~
 \frac{\sigma_n^2(\sigma_1^2+\sigma_n^2)}{\sigma_1^2\cos^2\tau+\sigma_n^2}
 \label{eq:mse_mmsedr_real}\\
 =&~
 \mbox{MSE}_{\rm MVDR} -
 \frac{\abs{c_1}^2\cos^2\tau}{\sigma_1^2\cos^2\tau+\sigma_n^2}
 \label{eq:mse_mmsedrmvdr_real}\\
 =&~
 \mbox{MSE}_{\rm ZF} -
 \frac{\sigma_n^4\tan^2\tau}{\sigma_1^2\cos^2\tau+\sigma_n^2}.
 \label{eq:mse_mmsedrzf_real}
\end{align}

\end{theorem}

The expression in \eqref{eq:mse_mvdr_real} implies
that, when $c_1\neq 0$
(in the presence of
temporal correlations between the desired and interfering sources),
the MSE of MVDR does not decrease monotonically in general
and may even increase as the noise power  $\sigma_n^2$ decreases.
This is also true for the case of many users as shown in Section \ref{sec:exp}.

\subsection{Complex Case for $J:=1$}

In an analogous way to the real case, the following theorems can be
verified.

\begin{theorem}
\label{theorem:rzf_complex_Nd}
Let $J:=1$, 
$\norm{\signal{h}_0}=\norm{\signal{h}_1}=1$,
$\innerprod{\signal{h}_0}{\signal{h}_1}=\sin \tau e^{i\phi_z}$,
the spatial-separation factor
$\tau\in[0,\pi/2)$,
$\phi_z\in[0,2\pi)$,
$\sigma_0^2>0$,
$\sigma_1^2>0$,
$\sigma_n^2>0$,
and 
the correlation factor
$c_1:=\abs{c_1}e^{i\phi_c}\in\{c\in \comp\mid \abs{c}\leq
 \sigma_0\sigma_1\}$, $\phi_c\in[0,2\pi)$.\footnote{The range $[0,\pi/2)$ of $\tau$
 is a half of the range $(-\pi/2,\pi/2)$ in Theorem \ref{theorem:rzf_real_Nd}
as the other half $(-\pi/2,0)$ is covered by letting $\phi_z:=\pi$.}
Here, the trivial case of $\sin\tau=1$ is excluded as in Theorem
 \ref{theorem:rzf_real_Nd}.
Then, the MSE is given,
as a function of the Lagrange multiplier $\lambda_\varepsilon$,
by
\begin{equation}
 {\rm MSE}(\lambda_\varepsilon) = \frac{\abs{\delta_2}^2(\sigma_1^2 \cos^2\tau +
  \sigma_n^2)}{g^2(\lambda_{\varepsilon})}
- \frac{2\sigma_n^2 \delta_1\tan\tau}{g(\lambda_{\varepsilon})}
+ \sigma_n^2(\tan^2\tau +1).
\label{eq:theorem_mse_complex}
\end{equation}
Here, $g(\lambda_{\varepsilon})$ is defined in \eqref{eq:def_g} and
\begin{align}
 \delta_1:=&~\sigma_n^2 \tan\tau -
 \abs{c_1}\cos\tau\cos(\phi_c+\phi_z), \label{eq:delta1}\\
 \delta_2:=&~\sigma_n^2 \tan\tau -
 \abs{c_1}\cos\tau\exp[i(\phi_c+\phi_z)]. \label{eq:delta2}
\end{align}

\begin{enumerate}
 \item Assume that $\delta_2=0$ ($\Rightarrow \delta_1=0$).
Then, the MSE reduces to
       $\sigma_n^2 (\tan^2\tau+1)$, which is constant in $\lambda_{\varepsilon}$.

 \item Assume that $\delta_2\neq0$.
\begin{enumerate}
 \item If 
\begin{equation}
 \gamma:=\dfrac{\delta_1\sigma_n^2\tan\tau}{\abs{\delta_2}^2} \leq 0~
(\Leftrightarrow \delta_1 \tan\tau\leq 0),
\end{equation}
${\rm MSE}(\lambda_\varepsilon)$ is monotonically decreasing, and
the ZF beamformer is optimal in the MSE sense.
 \item 
If $\gamma \geq 1$,
${\rm MSE}(\lambda_\varepsilon)$ is monotonically increasing, and thus is
       minimized by
$\lambda_\varepsilon:=0$.
\item If $\gamma \in (0,1)$,
 ${\rm MSE}(\lambda_\varepsilon)$  is minimized by
\begin{equation}
 \lambda_{\varepsilon}:=
\dfrac{\sigma_1^2\cos^2\tau + \sigma_n^2}{\cos^2\tau}
\left(\frac{1-\gamma}{\gamma}\right) >0.
\label{eq:theorem_lambda_comp}
\end{equation}

\end{enumerate}

\end{enumerate}

\end{theorem}

\begin{theorem}[Superiority of RZF]
\label{theorem:rzf_superiority}
Under the same settings as in Theorem \ref{theorem:rzf_complex_Nd},
the RZF beamformer with the optimal 
Lagrange multiplier
$\lambda_\varepsilon$
gives a strictly smaller MSE than
the MVDR and ZF beamformers (and achieves the minimum MSE under the distortionless constraint)
 if and only if $0<\delta_1\tan\tau<\abs{\delta_2}^2$.
Moreover, the same is true if (but not only if) 
$c_1\neq 0$ and
$\sin\tau\cos(\phi_c+\phi_z)<0$.
\end{theorem}

\begin{theorem}[Achieved MSE in complex case]
\label{theorem:complex_mse}
Assume the same settings as in Theorem \ref{theorem:rzf_complex_Nd}.
If $\delta_2(:=\sigma_n^2 \tan\tau - \abs{c_1}\cos\tau\exp[i(\phi_c+\phi_z)]) = 0$,
it holds that
$\mbox{MSE}_{\rm RZF}= \mbox{MSE}_{\rm ZF}=\mbox{MSE}_{\rm MVDR}$
for any $\lambda_\varepsilon\geq 0$.
Assume that $\delta_2\neq 0$.
Then, with the optimal 
Lagrange multiplier
$\lambda_\varepsilon$, it holds that
\begin{equation}
  \mbox{MSE}_{\rm RZF}= 
\begin{cases}
  \mbox{MSE}_{\rm ZF},
& \hspace*{-6em}\mbox{if } \gamma\left(:=\dfrac{\delta_1\sigma_n^2\tan\tau}{\abs{\delta_2}^2}\right)\leq 0,\\
  \mbox{MSE}_{\rm MMSE\mathchar`-DR} +
\left(1-\abs{\dfrac{\delta_1}{\delta_2}}^2\right)
 \dfrac{\sigma_n^4\tan^2\tau}{\sigma_1^2\cos^2\tau+\sigma_n^2}, & \mbox{if } \gamma\in(0,1),\\
  \mbox{MSE}_{\rm MVDR},
&  \mbox{if } \gamma\geq 1,
\end{cases}
\label{eq:MSE_complex}
\end{equation}
where the MSEs of the MVDR, ZF, and MMSE-DR beamformers are given by the
same expressions as in Theorem \ref{theorem:real_mse}.

\end{theorem}

\begin{remark}
[On Theorem \ref{theorem:complex_mse}: single-interference case]
\label{remark:theorem_complex_mse}
~
\begin{enumerate}
 \item 
Suppose that $\delta_1\neq 0$ ($\Rightarrow \delta_2\neq 0$),
 $\tau\in(0,\pi/2)$, $\sigma_n^2>0$, and $c_1\neq 0$.
Then, it can be shown that
MSE$_{\rm RZF}=$
MSE$_{\rm MMSE\mathchar`-DR}$ (i.e., RZF is optimal)
if and only if $\cos(\phi_c+\phi_z)=-1$
($\Leftrightarrow \delta_1=\delta_2$ and $\gamma\in(0,1)$).

 \item The ratio between ZF and MMSE-DR is given by
\begin{equation*}
\frac{ \mbox{MSE}_{\rm MMSE\mathchar`-DR}}{\mbox{MSE}_{\rm ZF}}=
1-\frac{1}{[1+(\sigma_1^2/\sigma_n^2)\cos^2\tau](1+\tan^{-2}\tau)},
\end{equation*}
which approaches zero as 
$\cos^2\tau\rightarrow 0$.

 \item 
The ratio between MVDR and MMSE-DR is given by
\begin{equation*}
\frac{ \mbox{MSE}_{\rm MMSE\mathchar`-DR}}{\mbox{MSE}_{\rm MVDR}}=
\frac{1}{1 +
\dfrac{\abs{c_1}^2\cos^2\tau}{\sigma_n^2(\sigma_1^2+\sigma_n^2)}}
\in\Bigg( \frac{\sigma_n^2}{\sigma_n^2+1},1 \Bigg],
\end{equation*}
where the lower bound corresponds to the extreme case of
       $\abs{c_1}^2=\sigma_1^2$
and $\cos^2 \tau=1$ in the limit of $\sigma_1^2\rightarrow
       \infty$, and this lower bound approaches zero as
$\sigma_n^2\rightarrow 0$.
The ratio increases as
$(\sigma_1^2/\sigma_n^2)\abs{c_1/\sigma_1}^2\cos^2\tau$ increases.

 \item The MSE of MVDR increases as the amplitude $\abs{c_1}$ of the
       correlation does, while those of MMSE-DR and ZF are independent
       of the correlation.
The performance of RZF depends on $\abs{c_1}$ in general
but is less sensitive than that of MVDR, as shown in Section
       \ref{sec:exp}.

 \item It holds that $\mbox{MSE}_{\rm RZF}=\mbox{MSE}_{\rm MVDR}$
when $c_1=0$ (there is no correlation between the desired
       and the interfering signals),
$\sigma_n^2\rightarrow +\infty$ (the noise power is huge), or
$\cos^2\tau=0$ ($\signal{h}_0$ and $\signal{h}_1$ are completely
       aligned).
This is consistent with Theorems \ref{theorem:rzf_real_Nd} and \ref{theorem:rzf_complex_Nd}.
In sharp contrast, 
$\mbox{MSE}_{\rm RZF}=\mbox{MSE}_{\rm ZF}$ when
$\sigma_n^2=0$ (the signal is noise free),
$\sin^2\tau=0$ ($\signal{h}_1$ is orthogonal to $\signal{h}_0$),
or $\sigma_1^2\rightarrow +\infty$ (the interference power is huge).

\end{enumerate}
\end{remark}

\subsection{Proofs}

The proofs of Theorems 1 -- 5 are given below.

\subsubsection{Proof of Theorem \ref{theorem:rzf_real_Nd}}

In the case of $J=1$,
\eqref{eq:jmse_distortionless} reduces to
\begin{equation}
 J_{\rm MSE}(\signal{w})=
\sigma_n^2 \norm{\signal{w}}^2+\sigma_1^2
|\signal{h}_1^{\sf H} \signal{w}|^2, ~\signal{w}\in C.
\label{eq:J_MSE_2dimcase}
\end{equation}
Now, applying the matrix inversion lemma to
\refeq{eq:wrzf_analytic_solution}, 
one can easily verify
that
$\norm{\signal{w}_{\rm RZF}}^2$ and
$\abs{\innerprod{\signal{h}_1}{\signal{w}_{\rm RZF}}}^2$ depend
only on the norms ($\norm{\signal{h}_0}$ and $\norm{\signal{h}_1}$)
and the inner product $\innerprod{\signal{h}_0}{\signal{h}_1}$.
This simple observation implies that 
the MSE of the RZF beamformer in the theorem can be expressed analytically
via analyzing the two-dimensional case\footnote{The
dimensionality reduction can also be understood through
the isomorphism between
${\rm span}\{\signal{h}_0,\signal{h}_1\}\subset \comp^N$ and $\comp^2$.
Indeed, the pair $(\signal{h}_0,\signal{h}_1)$ of vectors in Theorem
\ref{theorem:rzf_real_Nd} has
the same geometric structure as in Lemma \ref{lemma:rzf_real}.}
with such unit vectors
$\signal{h}_0\in\comp^2$ and $\signal{h}_1\in\comp^2$ that satisfy 
$\innerprod{\signal{h}_0}{\signal{h}_1}=\sin \tau$.
The proof is thus completed by proving the following lemma.

\begin{lemma}
\label{lemma:rzf_real}
Let $N:=2$, $J:=1$, $\signal{h}_0:=[0,1]^{\sf T}$,
 $\signal{h}_1:=[\cos \tau,\sin \tau]^{\sf T}$, $\tau\in(-\pi/2,\pi/2)$,
$\sigma_0^2>0$,
$\sigma_1^2>0$,
$\sigma_n^2>0$,
and $c_1\in[-\sigma_0\sigma_1,\sigma_0\sigma_1]$.
The RZF beamformer is then given by
\begin{equation}
\signal{w}_{\rm RZF}= \left[
\begin{array}{c}
-\dfrac {\cos \tau\left[(\sigma_1^2+\lambda_{\varepsilon})\sin\tau +
  c_1\right]}
 {(\sigma_1^2+\lambda_{\varepsilon}) \cos^2\tau + \sigma_n^2}
 \\
1
\end{array}
\right],
\label{eq:theorem_wrzf}
\end{equation}
for which the MSE is given by 
\eqref{eq:theorem_mse}.
\begin{enumerate}
 \item[(i)] Assume that $\delta=0$.
Then, MSE is given by
       $\sigma_n^2 (\tan^2\tau+1)$, which is constant in $\lambda_{\varepsilon}$.

 \item[(ii)] Assume that $\delta\neq0$.
If $\gamma:=\dfrac{\sigma_n^2\tan\tau}{\delta}\leq 0$,
${\rm MSE}(\lambda_\varepsilon)$ is monotonically decreasing, and 
the ZF beamformer (corresponding to the limit of
$\lambda_\varepsilon\rightarrow +\infty$) is optimal in the MSE sense.
If
$\gamma> 0$ ($\Rightarrow \sin\tau\neq 0$),
${\rm MSE}(\lambda_\varepsilon)$  is minimized by
\eqref{eq:theorem_lambda}.
Note here that $0<(\delta/(\sigma_n^2 \tan\tau)-1)/\cos^2 \tau = -c_1/(\sigma_n^2\sin\tau)$.
\end{enumerate}
\end{lemma}

{\it Proof:}
For the sake of simple notation, we let $c:=\cos\tau$ and $s:=\sin\tau$.
By 
$\signal{R}= \sigma_0^2\signal{h}_0\signal{h}_0^{\sf T}
 +\sigma_1^2\signal{h}_1\signal{h}_1^{\sf T} +
 c_1(\signal{h}_0\signal{h}_1^{\sf T} + \signal{h}_1\signal{h}_0^{\sf
 T}) +\sigma_n^2\signal{I}$,
we can verify that 
\begin{equation}
 \signal{R}_\varepsilon= \left[
\begin{array}{cc}
(\sigma_1^2 + \lambda_{\varepsilon})c^2+\sigma_n^2 & 
(\sigma_1^2 + \lambda_{\varepsilon})cs + c_1c \\
(\sigma_1^2 + \lambda_{\varepsilon})cs + c_1 c & 
~(\sigma_1^2 + \lambda_{\varepsilon})s^2+2c_1s+\sigma_n^2+\sigma_0^2 \\
\end{array}
\right],
\end{equation}
from which it follows that
\begin{equation}
 \signal{R}_\varepsilon^{-1}\signal{h}_0 \sim \left[
\begin{array}{c}
-c\left[
(\sigma_1^2 + \lambda_{\varepsilon})s+c_1
\right]\\
~(\sigma_1^2 + \lambda_{\varepsilon})c^2+\sigma_n^2 \\
\end{array}
\right].
\end{equation}
Hence, we obtain
\begin{equation}
 \signal{w}_{\rm RZF}=
\dfrac{\signal{R}_{\varepsilon}^{{\rm -1}}{\signal h}_{{\rm 0}}}
{\signal{h}_{{\rm 0}}^{\sf T}\signal{R}_{\varepsilon}^{{\rm
 -1}}\signal{h}_{{\rm 0}}}
= \left[
\begin{array}{c}
-\dfrac{c\left[
(\sigma_1^2 + \lambda_{\varepsilon})s+c_1\right]}
{(\sigma_1^2 + \lambda_{\varepsilon})c^2+\sigma_n^2}
 \\
1
\end{array}
\right].
\label{eq:w_rzf_2dimcase}
\end{equation}
Substituting \eqref{eq:w_rzf_2dimcase} into \eqref{eq:J_MSE_2dimcase}
yields \eqref{eq:theorem_mse}.

In the rest of the proof, we solely consider the case of 
$\delta\neq 0$, as the case of $\delta=0$ is clear.
In this case, 
define $x:=1/g(\lambda_\varepsilon)\in(0,1/(\sigma_1^2
 \cos^2\tau+\sigma_n^2)]$ for $\lambda_\varepsilon\in[0,+\infty)$.
Then, MSE can be seen as a quadratic function of $x$, and its minimum
 over $\real$ is achieved by
\begin{equation}
x^*:= 
\frac{1}{\sigma_1^2 \cos^2\tau+\sigma_n^2}\times
\frac{\sigma_n^2\tan\tau}{\underbrace{\sigma_n^2\tan\tau - c\cos\tau}_{=\delta}}.
\end{equation}
If $\gamma(:=(\sigma_n^2\tan\tau)/\delta)\leq 0$, 
$x^*\leq 0$ and hence
MSE is a monotonically increasing function of $x$
within the actual range $x\in(0,1/(\sigma_1^2
 \cos^2\tau+\sigma_n^2)]$.
If, on the other hand, 
 $\dfrac{\sigma_n^2\tan\tau}{\delta}\geq 1$, 
$x^*\geq 1/(\sigma_1^2
 \cos^2\tau+\sigma_n^2)$ and hence 
MSE is a monotonically decreasing function of $x$
within the range.
In this case, the minimum is clearly achieved by 
$x:= 1/(\sigma_1^2\cos^2\tau+\sigma_n^2) \Leftrightarrow
 \lambda_\varepsilon:=0$.
If  $\dfrac{\sigma_n^2\tan\tau}{\delta}\in(0,1)$, $x^*$ lies in the
 range, and its corresponding $\lambda_\varepsilon$ is given by
$\lambda_\varepsilon := -c_1\left[(\sigma_1^2/\sigma_n^2)\cos^2\tau + 1\right]/\sin\tau$.
\migip

\subsubsection{Proof of Theorem \ref{theorem:real_mse}}
\label{subsubsec:proof_Theorem2}

We first prove the following proposition.
\begin{proposition}
\label{proposition:mmsedr}
Under the same settings as Lemma \ref{lemma:rzf_real}, the MMSE
 beamformer under $\signal{w}\in C$ are given respectively by
 \begin{align}
 \signal{w}_{\rm MMSE\mathchar`-DR}
=&~
 \left[
\begin{array}{c}
-\dfrac {\sigma_1^2 \sin\tau\cos\tau}
 {\sigma_1^2\cos^2\tau + \sigma_n^2}
 \\
1
\end{array}
\right],
\label{eq:mmsedr_2dim_real}\\
 \signal{w}_{\rm MVDR}
=&~
 \left[
\begin{array}{c}
-\dfrac{\cos\tau (\sigma_1^2 \sin\tau+ c_1)}
 {\sigma_1^2\cos^2\tau + \sigma_n^2}
 \\
1
\end{array}
\right]
\nonumber\\
= &~
 \signal{w}_{\rm MMSE\mathchar`-DR}-
 \left[
\begin{array}{c}
\dfrac {c_1 \cos\tau}
 {\sigma_1^2\cos^2\tau + \sigma_n^2}
 \\
0
\end{array}
\right].
\label{eq:mvdr_2dim_real}
\end{align}
\end{proposition}

{\it Proof:}
 Equation \eqref{eq:mmsedr_2dim_real} can be verified by substituting the following equations into
 \eqref{eq:mmsedr}:
\begin{align}
 \widetilde{\signal{R}} =& ~\signal{Q} \signal{\Lambda} \signal{Q}^{\sf T},\nonumber\\
\signal{h}_0 =&~ \signal{Q}\signal{h}_1^{\perp},
\end{align}
where
\begin{align}
\signal{Q}:=&~\left[\signal{h}_1 ~ \signal{h}_1^{\perp}\right]:=
\left[\begin{array}{cc}
 \cos\tau &\sin\tau  \\
\sin\tau & - \cos\tau
  \end{array}\right],
\nonumber\\
\signal{\Lambda}:=&~
\left[\begin{array}{cc}
 \sigma_1^2 + \sigma_n^2 & 0  \\
0 & \sigma_n^2
  \end{array}\right].
\end{align}
Note here that the matrix $\signal{Q}$ is symmetric; i.e., 
$\signal{Q}^{\sf T}=\signal{Q}$.
Equation \eqref{eq:mvdr_2dim_real} can readily be verified by substituting
$\lambda_\varepsilon:=0$ into \eqref{eq:w_rzf_2dimcase}.
\migip

\noindent{\it Proof  of Theorem \ref{theorem:real_mse}:}
Substituting \eqref{eq:mmsedr_2dim_real} and \eqref{eq:mvdr_2dim_real} into 
\refeq{eq:J_MSE_2dimcase} yields \eqref{eq:mse_mvdr_real}
and \eqref{eq:mse_mmsedr_real}.
The MSE \eqref{eq:mse_zf_real} of ZF is verified by considering the
limit of $\lambda_\varepsilon\rightarrow +\infty$ in \eqref{eq:theorem_mse}.
The rest to prove is \eqref{eq:mse_rzf_real}, in which the cases of 
$\gamma\leq 0$ and $\gamma\geq 1$ can be verified straightforwardly from
Theorem \ref{theorem:rzf_real_Nd}.
The case of $\gamma\in (0,1)$ can be verified by substituting 
\eqref{eq:def_delta} -- \eqref{eq:theorem_lambda}
into \eqref{eq:theorem_mse}.


\begin{figure}[t!]
\centering
\psfrag{w1}[Bl][Bl][0.9]{$w_1$}
\psfrag{w2}[Bl][Bl][0.9]{$w_2$}
 \subfigure[$c_1:=0.99$, $\tau:=\pi/6$ ($\gamma=-$2.0619)]{
\hspace*{-1.5em}\includegraphics[width =.7\textwidth]{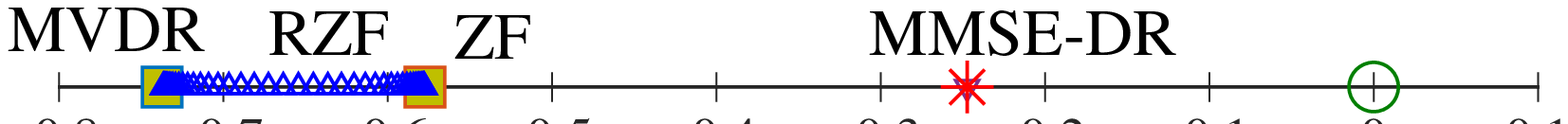}
}
	\centering
 \subfigure[$c_1:=-0.2$, $\tau:=\pi/6$ ($\gamma=$0.7692)]{
\hspace*{-1.5em}\includegraphics[width =.7\textwidth]{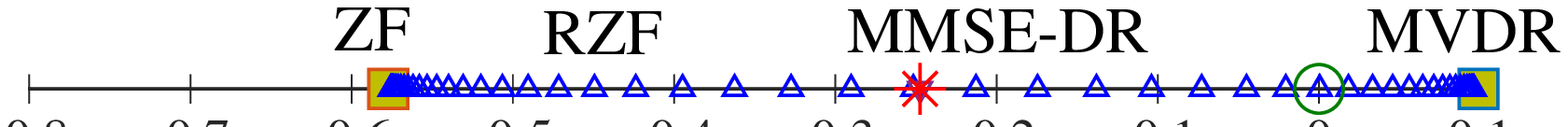}
 }

 \subfigure[$c_1:=0.1$, $\tau:=\pi/6$ ($\gamma=$1.1765)]{
\hspace*{-1.5em}\includegraphics[width =.7\textwidth]{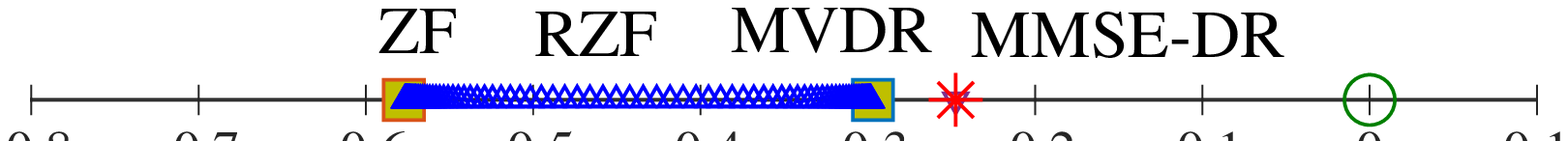}
 }

	\caption{The $w_1$ coordinate for $N:=2$ and $J:=1$ with
 $\signal{h}_0:=[0,1]^{\sf T}$
 under  $\sigma_1^2 := \sigma_n^2 := 1.0$.
For RZF, $\lambda_\varepsilon:=10^\alpha$, where $\alpha$ changes from
 $-2$ to $3$ at interval $0.1$.}
	\label{fig:1jam}
\end{figure}

\begin{figure}[t!]
\psfrag{ZF}[Bl][Bl][0.9]{ZF}
\psfrag{RZF}[Bl][Bl][0.9]{RZF}
\psfrag{MVDR}[Bl][Bl][0.9]{MVDR}
\psfrag{MMSE-DR}[Bl][Bl][0.9]{MMSE-DR}
\psfrag{w1}[Bl][Bl][0.9]{$w_1$}
\psfrag{w2}[Bl][Bl][0.9]{$w_2$}
\psfrag{h0}[Bl][Bl][0.9]{$\signal{h}_0$}
\psfrag{o}[Bl][Bl][0.9]{$\signal{0}$}
\psfrag{C}[Bl][Bl][0.9]{$C$}
	\centering
 \subfigure[]{
\includegraphics[width =.7\textwidth]{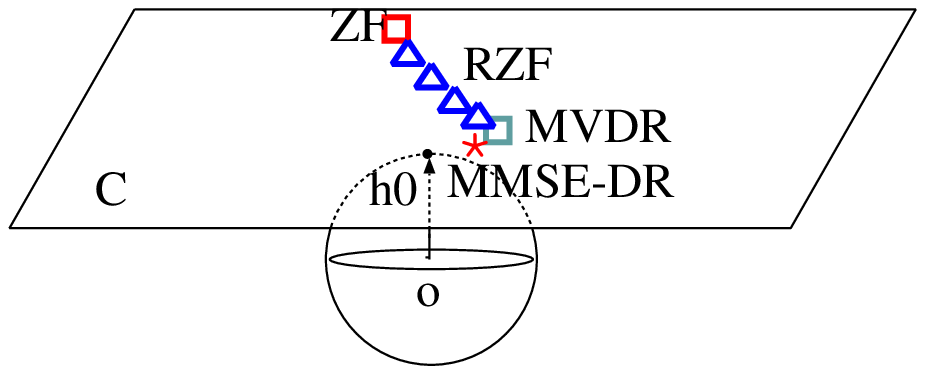}
 }
\centering
 \subfigure[]{
\psfrag{MMSE-DR}[Bl][Bl][.9]{\hspace*{1em}MMSE-DR}
\includegraphics[width =.7\textwidth]{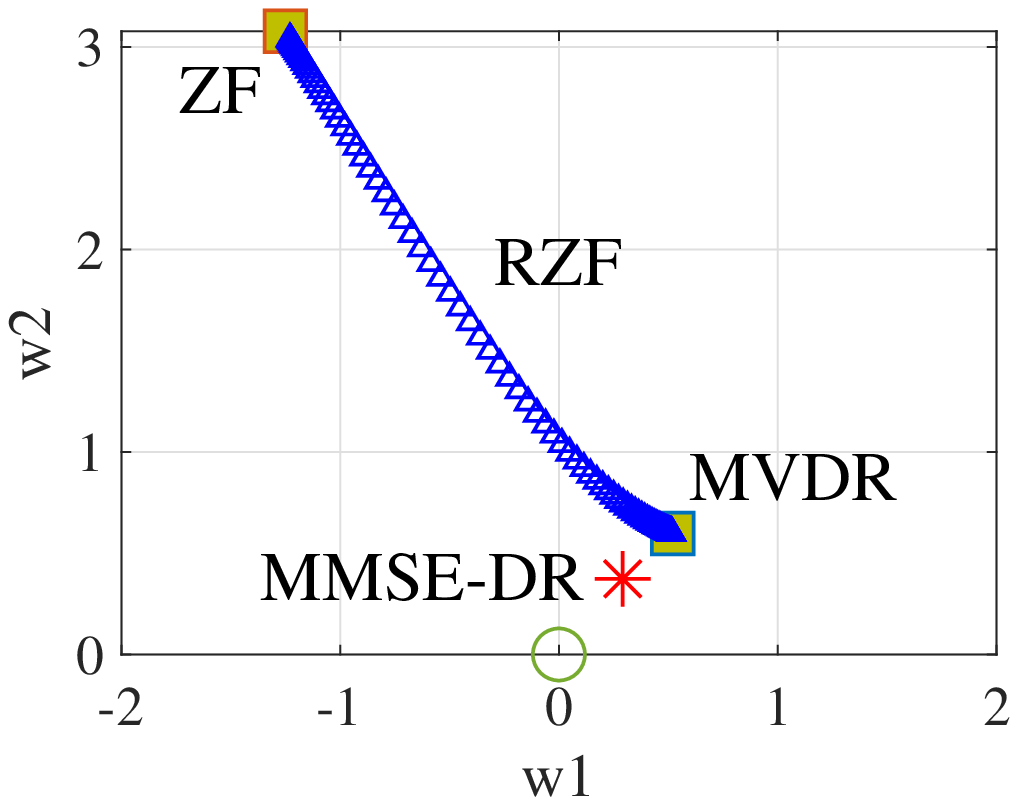}
}

	\caption{A possible situation in two-interference case
 with $\signal{h}_0:=[0,0,1]^{\sf T}$  ($N:=3$, $J:=2$):
(a) a schematic illustration and 
(b) the projections of the beamformers on $C$ onto
the $w_1$-$w_2$ plane for
 $\sigma_1^2:=\sigma_n^2:=1.0$,
$c_1:=c_2:=c_{1,2}:=0.6$, 
$\signal{h}_1:=[0,\cos(3\pi/5),\sin(3\pi/5)]$, and
$\signal{h}_2:=[\cos(5\pi/6),\sin(5\pi/6)\cos(5\pi/6),\sin(5\pi/6)\sin(5\pi/6)]$.
For RZF, $\lambda_\varepsilon:=10^\alpha$ with $\alpha$ changing from
 $-2$ to $3$ at interval $0.05$.}
	\label{fig:3d}
\end{figure}


\begin{remark}
\label{remark:noiseamp_real}
Figure \ref{fig:1jam} illustrates
the three cases of $\gamma$ in Theorem \ref{theorem:rzf_real_Nd} for $N:=2$:
(a) $\gamma\leq 0$, 
(b) $\gamma\in (0,1)$, and
(c) $\gamma\geq 1$.
If $c_1:=0$ for $J:=1$,
the $w_1$ components of the MVDR, ZF, RZF, and MMSE-DR beamformers have
 the same sign as $-\sin\tau$; note here that $\cos\tau>0$.
This is clear from the nulling constraint for ZF;
see Lemma  \ref{lemma:rzf_real} and Proposition \ref{proposition:mmsedr}
for the other beamformers.
If $c_1\neq 0$, on the other hand, 
the MVDR, RZF, and MMSE-DR beamformers may have the opposite sign from 
$-\sin\tau$ (see Figure \ref{fig:1jam}(b)),
while only the ZF beamformer sticks to the same sign as $-\sin\tau$.
This means that RZF may have a smaller norm than MVDR 
(and also than ZF), implying that it may suppress the noise
better than MVDR.
This also applies to the case of $J\geq 2$ as illustrated in
Figure \ref{fig:3d}, in which some of the RZF beamformers
have a smaller norm than MVDR.
The MSE expression given in \eqref{eq:jmse_distortionless} implies that
the MSE (under the distortionless constraint) depends on the norm
 $\norm{\signal{w}}^2$ of the beamformer and
 the interference leakage $\innerprod{\signal{w}}{\signal{h}_j}$.
In the particular case of $J:=1$, 
MSE is given by the simple sum of
 $\sigma_n^2\norm{\signal{w}}^2$ and $\sigma_1^2\abs{\innerprod{\signal{w}}{\signal{h}_1}}^2$.
To attain a small MSE (under the distortionless constraint),
$\signal{w}$ needs to be as close as possible to the origin, and
orthogonal to $\signal{h}_1$ as much as possible, simultaneously.
The optimal beamformer has the best balance between those two terms, while ZF over-weights
the second one.
Indeed, RZF may attain better noise-and-interference suppression
performance than MVDR for multiple interfering sources, as shown in Section \ref{sec:exp}.
\end{remark}

\subsubsection{Proof of Theorem \ref{theorem:rzf_complex_Nd}}

The claim can be verified by proving the following lemma
due to the same arguments as given
in the proof of Theorem \ref{theorem:rzf_real_Nd}.

\begin{lemma}
\label{lemma:rzf_complex}
Let $N:=2$, $J:=1$, $\signal{h}_0:=[0,1]^{\sf T}$,
 $\signal{h}_1:=[\cos \tau, z]^{\sf T}:=[\cos \tau,\sin\tau e^{i\phi_z}]^{\sf T}$
with $\tau\in[0,\pi/2)$ and $\phi_z\in[0,2\pi)$,
$\sigma_0^2>0$,
$\sigma_1^2>0$,
$\sigma_n^2>0$,
and $c_1:=\abs{c_1}e^{i\phi_c}\in\{c\in \comp\mid \abs{c}\leq
\sigma_0 \sigma_1\}$, $\phi_c\in[0,2\pi)$.\footnote{One may consider the more
 general case of $\cos \tau e^{i\phi}$ for some $\phi\in[0,2\pi)$,
 instead of $\cos \tau$ of zero phase.
However, the results in the theorem
(excluding the expression of $\signal{w}_{\rm RZF}$)
are independent of the choice of $\phi$,
and we can thus assume that $\phi:=0$ without any essential loss of generality.}
The RZF beamformer is then given by
\begin{equation}
\signal{w}_{\rm RZF}= \left[
\begin{array}{c}
-\dfrac {\cos \tau\left[(\sigma_1^2+\lambda_{\varepsilon})z^* +
  c_1\right]}
 {(\sigma_1^2+\lambda_{\varepsilon}) \cos^2\tau + \sigma_n^2}
 \\
1
\end{array}
\right],
\label{eq:theorem_wrzf}
\end{equation}
for which the MSE is given by
\eqref{eq:theorem_mse_complex}.

\begin{enumerate}
 \item[(i)] Assume that $\delta_2=0$ ($\Rightarrow \delta_1=0$), where
$\delta_1$ and $\delta_2$ are defined in
\eqref{eq:delta1} and  \eqref{eq:delta2}, respectively.
Then, the MSE is given by
       $\sigma_n^2 (\tan^2\tau+1)$, which is constant in $\lambda_{\varepsilon}$.

 \item[(ii)] Assume that $\delta_2\neq0$.
\begin{enumerate}
 \item If $\gamma:=\dfrac{\delta_1\sigma_n^2\tan\tau}{\abs{\delta_2}^2} \leq 0$
($\Leftrightarrow \delta_1 \tan\tau\leq 0$),
${\rm MSE}(\lambda_\varepsilon)$ is monotonically decreasing, and
the ZF beamformer is optimal in the MSE sense.
 \item 
If $\gamma \geq 1$,
${\rm MSE}(\lambda_\varepsilon)$ is monotonically increasing, and thus is
       minimized by
$\lambda_\varepsilon:=0$.
\item If $\gamma \in (0,1)$,
 ${\rm MSE}(\lambda_\varepsilon)$  is minimized by
\eqref{eq:theorem_lambda_comp}.
\end{enumerate}

\end{enumerate}

\end{lemma}

{\it Proof:}
The claim can be verified in an analogous way to the proof of Lemma \ref{lemma:rzf_real}.
\migip

\subsubsection{Proof of Theorem \ref{theorem:rzf_superiority}}

The former part is clear from the proof of Lemma \ref{lemma:rzf_real};
note that the necessary and sufficient condition
 $0<\delta_1\tan\tau<\abs{\delta_2}^2$ implies
$\delta_2\neq 0$ and 
 $\dfrac{\delta_1\sigma_n^2\tan\tau}{\abs{\delta_2}^2} \in (0,1)$.
To verify 
the latter part (the sufficient condition),
we assume that
$c_1\neq 0$ and
$\sin\tau\cos(\phi_c+\phi_z)<0$.
Then, we immediately have
\begin{align}
 \delta_1\tan\tau =&~ (\sigma_n^2 \tan\tau -
 \abs{c_1}\cos\tau\cos(\phi_c+\phi_z))\tan\tau
\nonumber\\
=&~\sigma_n^2 \tan^2\tau -
 \abs{c_1}\sin\tau\cos(\phi_c+\phi_z)>0.
\end{align}
Observing that
\begin{equation}
\abs{\delta_2}^2=
\delta_1^2 + 
 (\abs{c_1}\cos\tau\sin(\phi_c+\phi_z))^2,
\end{equation}
we can verify that
\begin{align}
\hspace*{-1.5em}&0<  -\abs{c_1}\cos^2\tau\frac{\cos(\phi_c+\phi_z)}{\sin\tau}
+ \frac{ (\abs{c_1}\cos\tau\sin(\phi_c+\phi_z))^2}{\delta_1\tan\tau}
\nonumber\\
\hspace*{-2em}&\Leftrightarrow
\sigma_n^2< 
\frac{\delta_1}{\tan\tau}
+ \frac{ (\abs{c_1}\cos\tau\sin(\phi_c+\phi_z))^2}{\delta_1\tan\tau}
\nonumber\\
\hspace*{-2em}&\Leftrightarrow
\sigma_n^2< 
\frac{\abs{\delta_2}^2}{\delta_1\tan\tau},
\end{align}
from which it follows readily that 
$\delta_2\neq 0$ and 
 $\dfrac{\delta_1\sigma_n^2\tan\tau}{\abs{\delta_2}^2}<1$.

\subsubsection{Proof of Theorem \ref{theorem:complex_mse}}

In an analogous way to the proof of Theorem \ref{theorem:real_mse},
the claim is verified with the following proposition.

\begin{proposition}
\label{proposition:mmsedr_complex}
Under the same settings as in Lemma \ref{lemma:rzf_complex}, 
the MMSE-DR and the MVDR beamformers are given respectively by
 \begin{align}
 \signal{w}_{\rm MMSE\mathchar`-DR}
=&~
 \left[
\begin{array}{c}
-\dfrac {\sigma_1^2 z^* \cos\tau}
 {\sigma_1^2\cos^2\tau + \sigma_n^2}
 \\
1
\end{array}
\right],
\label{eq:mmsedr_2dim}\\
 \signal{w}_{\rm MVDR}
=&~
 \left[
\begin{array}{c}
-\dfrac{\cos\tau (\sigma_1^2 z^* + c_1)}
 {\sigma_1^2\cos^2\tau + \sigma_n^2}
 \\
1
\end{array}
\right]
\nonumber\\
= &~
 \signal{w}_{\rm MMSE\mathchar`-DR}-
 \left[
\begin{array}{c}
\dfrac {c_1 \cos\tau}
 {\sigma_1^2\cos^2\tau + \sigma_n^2}
 \\
0
\end{array}
\right].
\label{eq:mvdr_2dim}
\end{align}
\end{proposition}

{\it Proof:}
 The claim can be verified in an analogous way to the proof of
 Proposition
 \ref{proposition:mmsedr}.
\migip

\begin{remark}
\label{remark:noiseamp_complex}
In the case of $c_1:=0$, 
the $w_1$ components of the MVDR, ZF, RZF, and MMSE-DR beamformers have
the same ``phase'' as $-\cos\tau z^*$ (see Lemma  \ref{lemma:rzf_complex} and Proposition
 \ref{proposition:mmsedr_complex}; cf.~Remark \ref{remark:noiseamp_real}).
If $c_1\neq 0$, on the other hand, 
the MVDR, RZF, and MMSE-DR beamformers have different phases in general from 
$-\cos\tau z^*$,
while the ZF beamformer sticks to the same phase as $-\cos\tau z^*$.
Hence, RZF, which resides between ZF and MVDR, possibly has a strictly smaller norm
than those two ends, as in the real case.
This implies that RZF has a potential to suppress the noise, as well as the interference,
more efficiently than MVDR,
because the impact of noise on the MSE is proportional to $\norm{\signal{w}}^2$
(see \eqref{eq:jmse_distortionless} in Lemma \ref{lemma:mse}).
Compared to ZF, RZF with an appropriate $\lambda_\varepsilon$ can achieve
a significantly better noise suppression performance
while suppressing the interference at the same level approximately, as
shown in Section \ref{sec:exp}.
\end{remark}

 \begin{figure}[t!]
 \psfrag{c}[Bl][Bl][0.9]{$\rho_j$}
 		\centering
\begin{tabular}{cc}
\hspace*{-1.5em}
 \subfigure[ $N=16$, $J=7$]{
\includegraphics[width =0.52\textwidth]{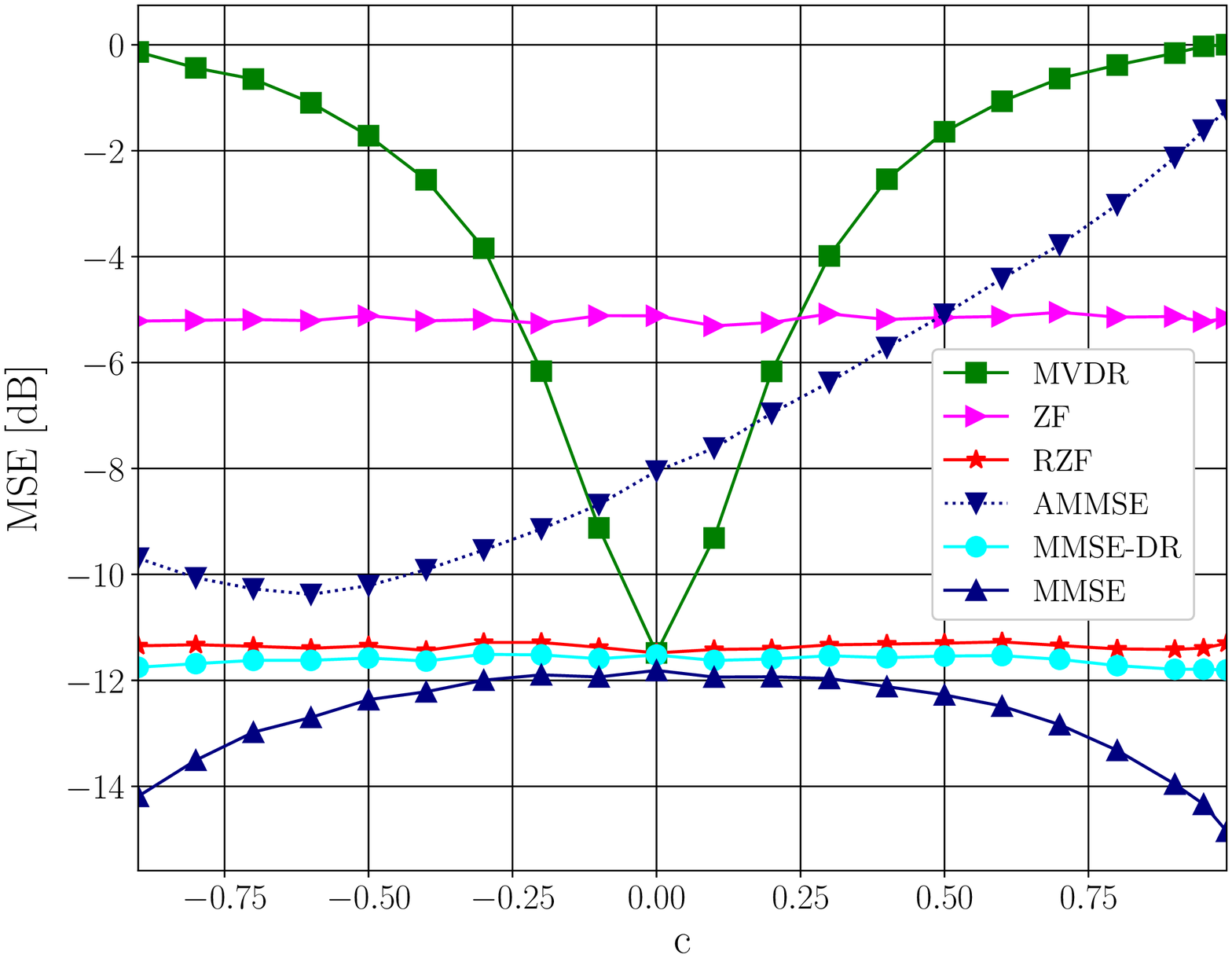}} &
\hspace*{-2em}
 \subfigure[ $N=64$, $J=19$]{
\includegraphics[width =0.52\textwidth]{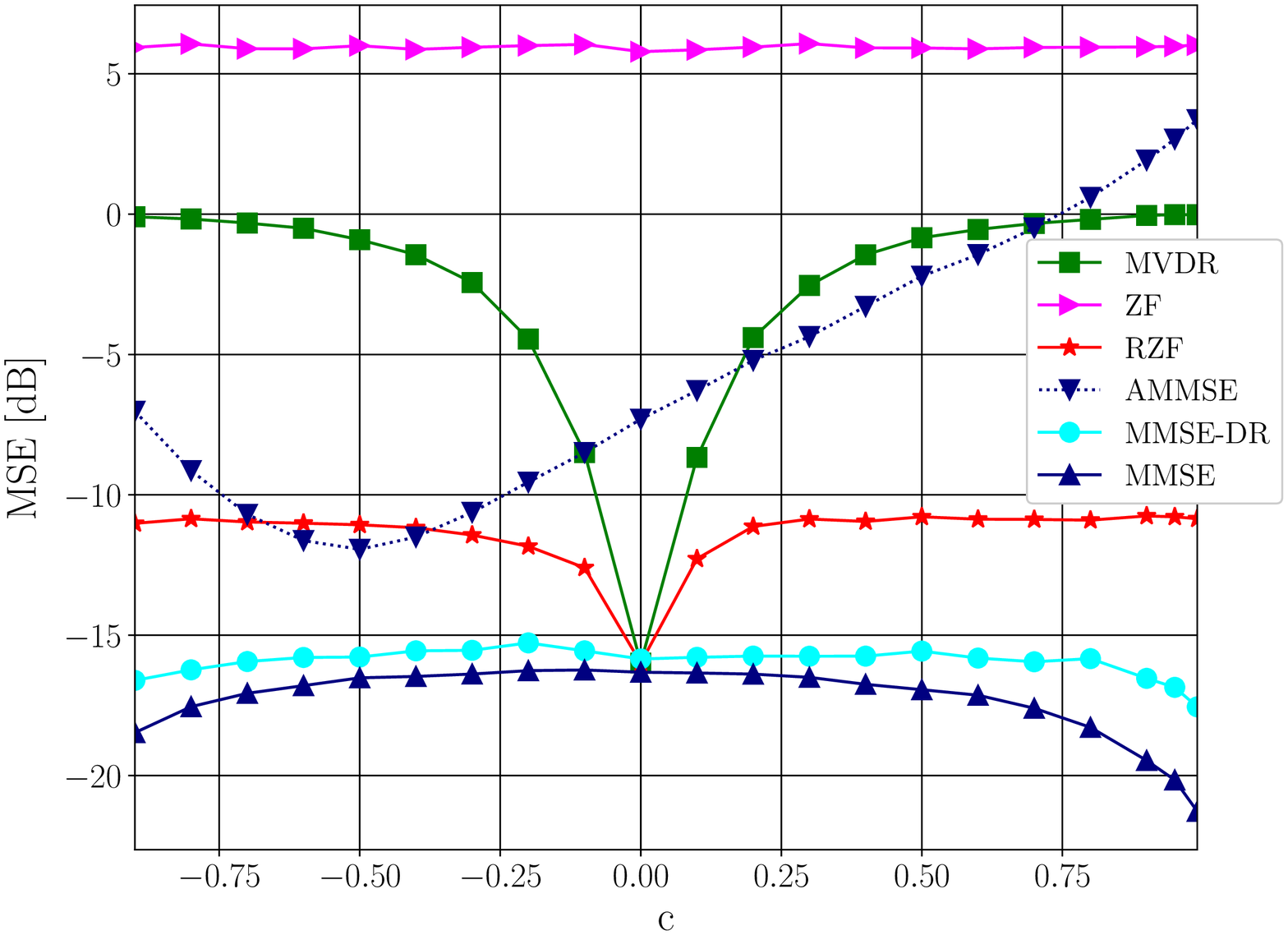}}
\end{tabular}
 \vspace{-5pt}
 	\caption{Performances across
  $\rho_j$ for $\beta:=0.8$,  $\varepsilon_\rho:=0.1$,
 and $\varepsilon_\phi:=\pi/12$
 under SNR 0 dB and SIR 0 dB. }
 	\label{fig:change_rhoj}
 \end{figure}

 \begin{figure}[t!]
 \psfrag{e}[Bl][Bl][0.9]{$\varepsilon$}
 		\centering
\begin{tabular}{cc}
\hspace*{-1.5em}
 \subfigure[ $N=16$, $J=7$]{
\includegraphics[width = 0.52\textwidth]{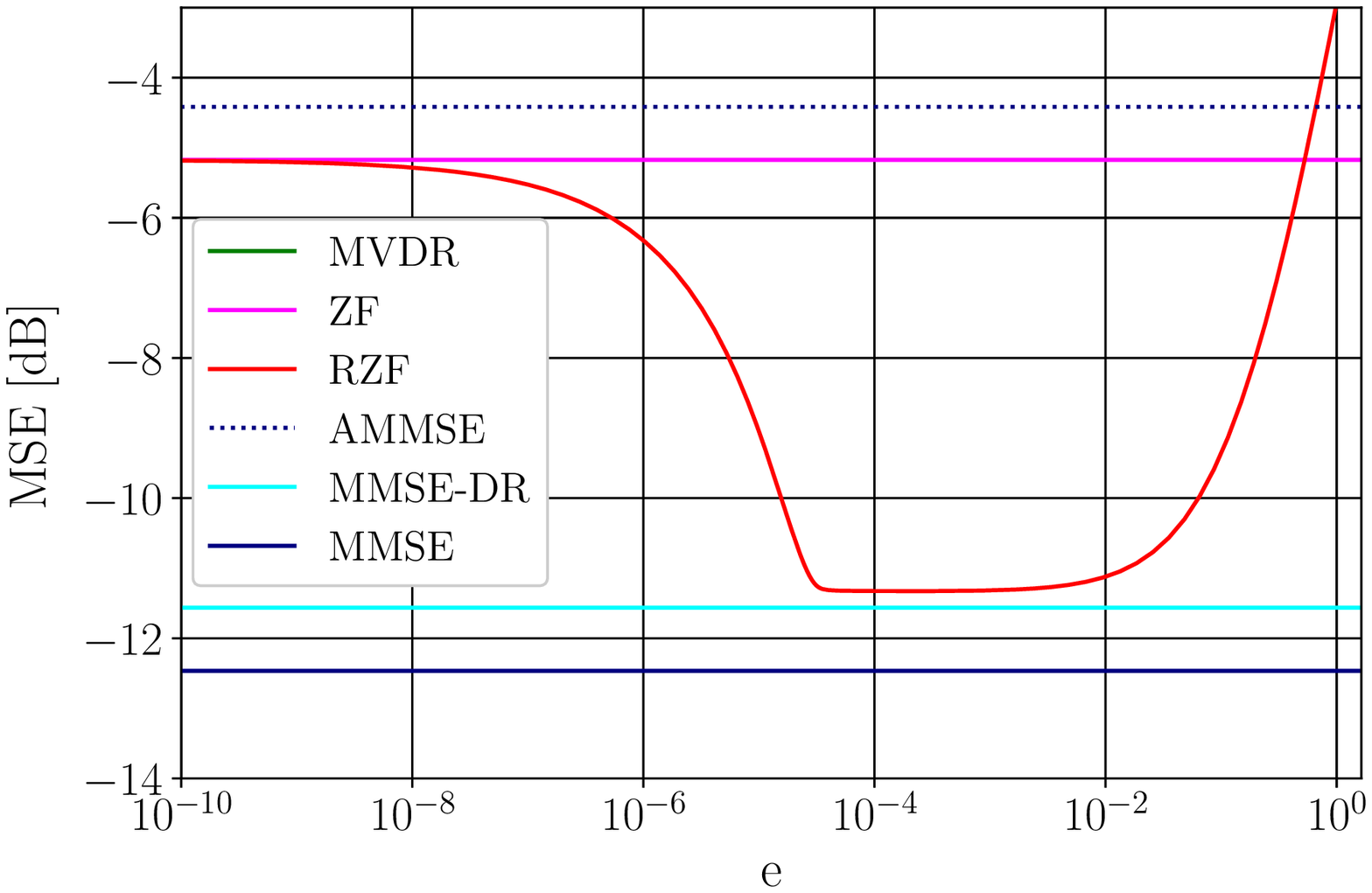}} &
\hspace*{-2em}
 \subfigure[ $N=64$, $J=19$]{
\includegraphics[width =0.52\textwidth]{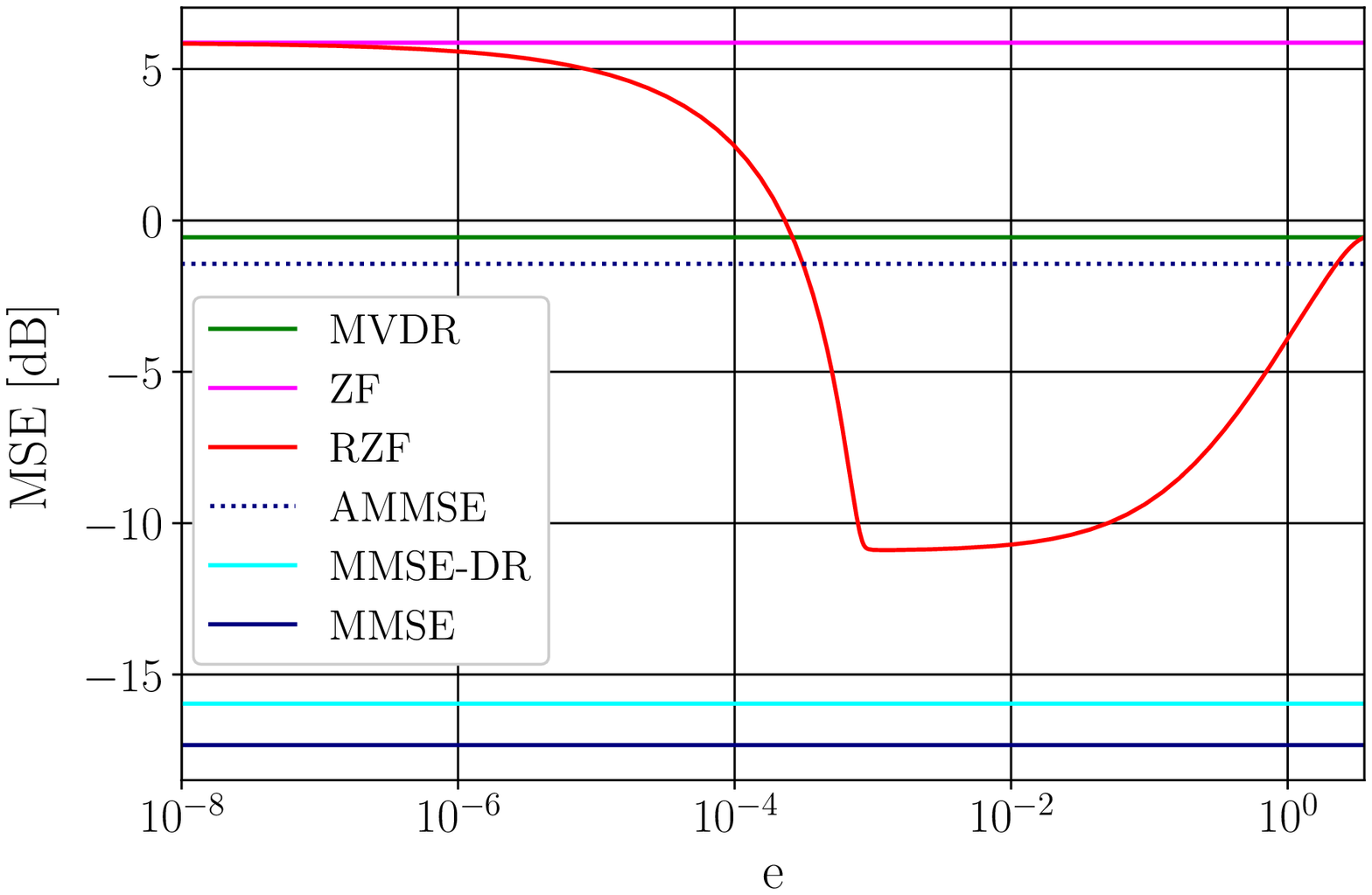}}
\end{tabular}
 \vspace{-5pt}
 	\caption{Sensitivity of RZF to the parameter
 $\varepsilon$ 
($\rho_j:=0.6$, $\beta:=0.8$, $\varepsilon_\rho:=0.1$, $\varepsilon_\phi:=\pi/12$, SNR 0 dB, SIR 0 dB).}
 	\label{fig:rzf}
 \end{figure}

 \begin{figure}[t!]
 		\centering
 		\vspace*{5pt}
 \psfrag{b}[Bl][Bl][0.9]{$\beta$}
 \psfrag{c}[Bl][Bl][0.9]{$\varepsilon_\rho$}
\begin{tabular}{cc}
\hspace*{-1.5em}
 \subfigure[]{
\includegraphics[width = 0.52\textwidth]{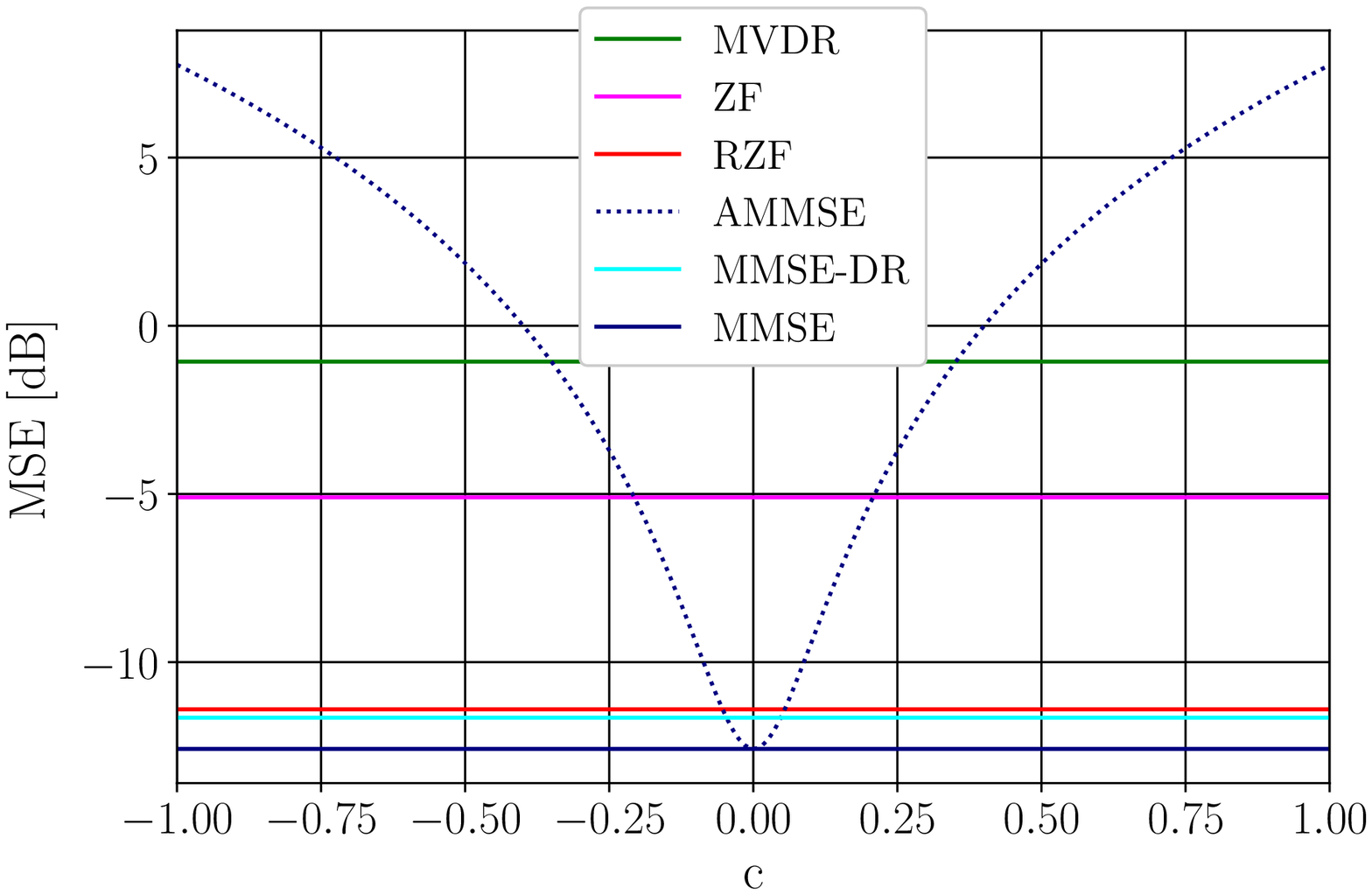}} &
\hspace*{-2em}
 \subfigure[]{
\includegraphics[width =0.52\textwidth]{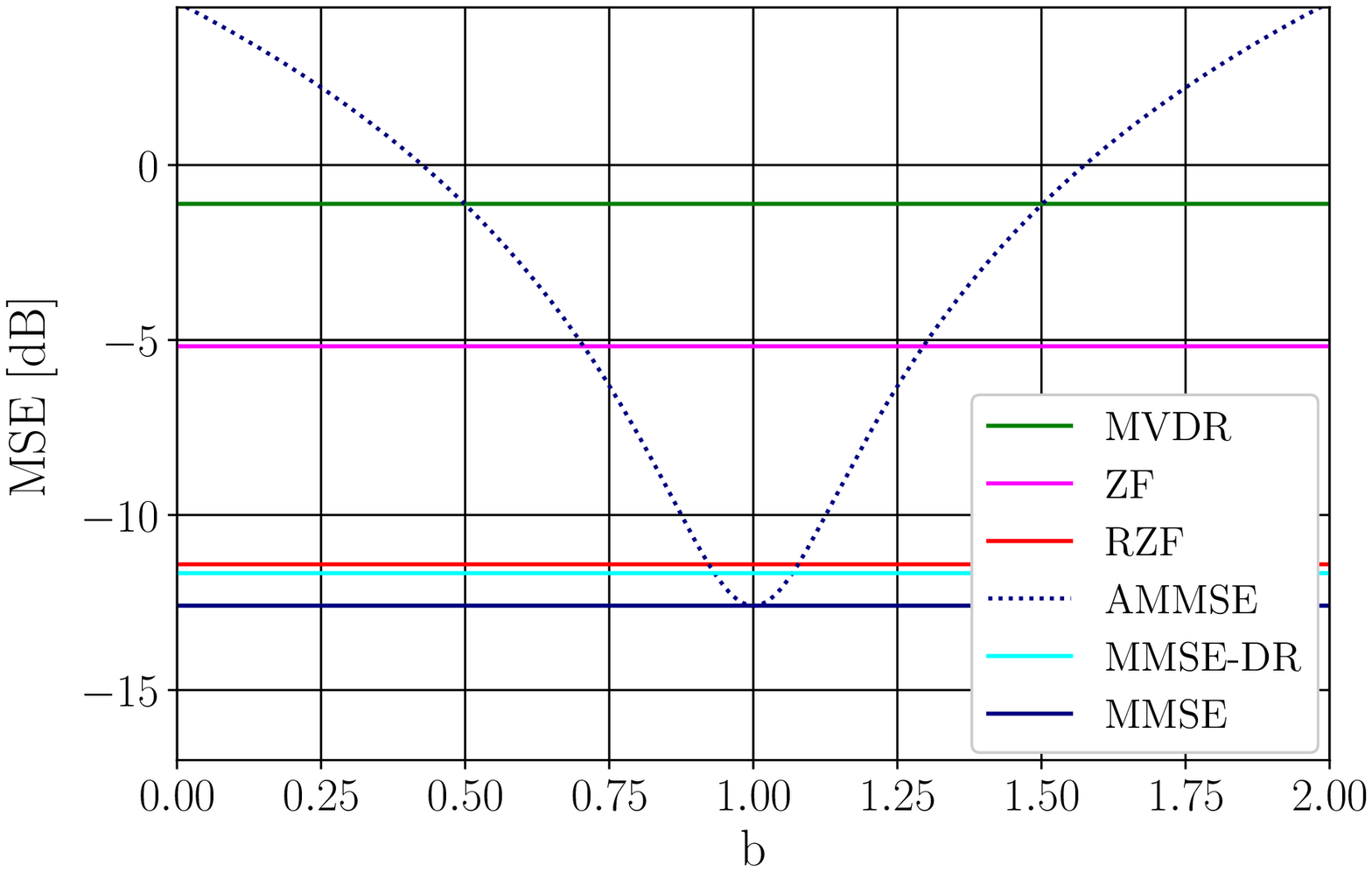}}
\end{tabular}

 \vspace{-5pt}
 	\caption{Sensitivity of the A-MMSE beamformer
to the errors $\varepsilon_\rho$ and $\beta$ ($\rho_j:=0.6$, SNR 0 dB, SIR 0 dB).}
 	\label{fig:approximate_mmse}
 \end{figure}

\section{Simulation Study of RZF Beamformer for Multiple Interference Case}
\label{sec:exp}

The basic performance of the RZF beamformer is studied
with toy models, and its efficacy is then shown
in application to brain activity reconstruction with EEG measurements.
We define SNR and the signal to interference ratio (SIR)
as the ratios of the power of desired signal projected onto
sensors to the power of noise and interference counterparts,
respectively; i.e.,
SNR
$:=\sum_{i=1}^{N}E(\abs{h_{0,i}s_0(k)}^2)/\sum_{i=1}^{N}E(\abs{n_i(k)}^2)$,
and 
${\rm SIR}:=\sum_{i=1}^{N}E(\abs{h_{0,i}s_0(k)}^2)/$
$\sum_{i=1}^{N}E(|\sum_{j=1}^{J}h_{j,i}s_j(k)|^2)$.
The parameter $\epsilon$ is tuned manually by grid search
for each setting in all simulations but the one which studies the
sensitivity of RZF to the value of $\epsilon$
to show the potential performance of RZF.

\subsection{Basic Performance with Toy Model}
\label{subsec:communication}

The uniform linear array is considered with
the array response 
$\signal{h}_j:= \signal{h}(\theta_j):=
\frac{1}{\sqrt{N}}[1,e^{2\pi i\frac{d}{\lambda} \cos\theta_j},
\cdots,e^{2\pi i (N-1) \frac{d}{\lambda}\cos\theta_j}]^{\sf T}\in\comp^N$
at the receiver, where 
$\theta_j\in[0,\pi]$ is the direction of arrival (DOA) of
the $j$th signal, 
$\lambda$ is the carrier wavelength, and
$d:=\lambda/2$ is the antenna spacing
\cite{trees02_book}. 
Two scenarios are considered regarding the array size:
(a) $N=16$ and $J=7$ (middle array size), and
(b) $N=64$ and $J=19$ (large array size).
In each setting, we let
$\{\theta_j\}_{j=0}^J = \{(j+1)\pi/(J+2)\}_{j=0}^J$
with $\theta_0 := \lceil (J+1)/3 \rceil\pi/(J+2)$.
The desired signal is drawn from 
the i.i.d.~standard complex Gaussian distribution, i.e.,
$s_0(k)\sim \mathcal{CN}(0,1)$.
For convenience in controlling the correlation between $s_0(k)$ and
$s_j(k)$,
the interfering signals are generated as
\begin{equation}
 s_j(k):=\dfrac{\sigma_j e^{i\phi_j}}{\sqrt{1+\sigma_v^2}} (s_0(k) +
  v_j(k)),
\label{eq:sjk_generation}
\end{equation}
where
$\sigma_j^2:=E(\abs{s_j(k)}^2)$, $v_j(k)\sim \mathcal{CN}(0,\sigma_v^2)$,
and the relative phase of the $j$th source is set to $\phi_j:=0$.
In this case, the correlation coefficient between $s_0(k)$ and $s_j(k)$
is given by
\begin{equation}
\rho_j:=c_j/(\sigma_0 \sigma_j) = e^{i\phi_j}/\sqrt{1 +
 \sigma_v^2}\in\{c\in\comp\mid \abs{c}\in(0,1]\}.
\end{equation}
The desired-signal power is fixed to $\sigma_0^2:=1$, and the powers
of the interference and noise are changed
according to the SIR and SNR, respectively.


 \begin{figure}[t!]
 		\centering
 		\vspace*{5pt}
\begin{tabular}{cc}
\hspace*{-1.5em}
 \subfigure[$N=16$, $J=7$, SIR $=$ 0 dB]{
\includegraphics[width = 0.52\textwidth]{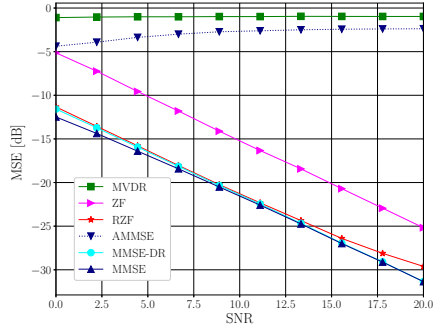}} &
\hspace*{-2em}
 \subfigure[$N=64$, $J=19$, , SIR $=$ 0 dB]{
\includegraphics[width =0.52\textwidth]{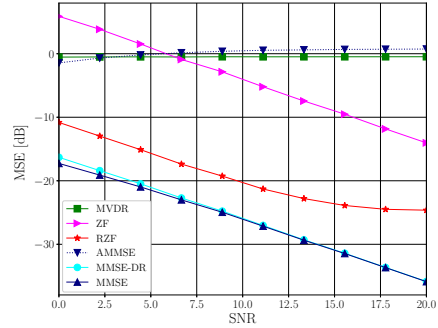}}
\end{tabular}

\begin{tabular}{cc}
\hspace*{-1.5em}
 \subfigure[$N=16$, $J=7$, SIR $=$ 10 dB]{
\includegraphics[width = 0.52\textwidth]{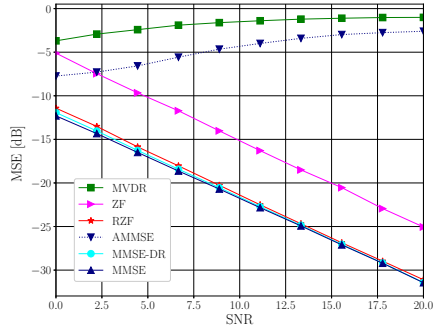}} &
\hspace*{-2em}
 \subfigure[$N=64$, $J=19$, SIR $=$ 10 dB]{
\includegraphics[width =0.52\textwidth]{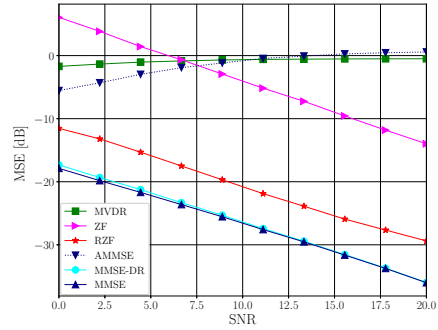}}
\end{tabular}

 \vspace{-5pt}
 	\caption{Performances  across SNR 
  for $\rho_j:=0.6$, $\beta:=0.8$, $\varepsilon_\rho:=0.1$, $\varepsilon_\phi:=\pi/12$.}
 	\label{fig:different_snr}
 \end{figure}

 \begin{figure}[t!]
 		\centering
 		\vspace*{5pt}
\begin{tabular}{cc}
\hspace*{-1.5em}
 \subfigure[$N=16$, $J=7$, SNR $=$ 0 dB]{
\includegraphics[width = 0.52\textwidth]{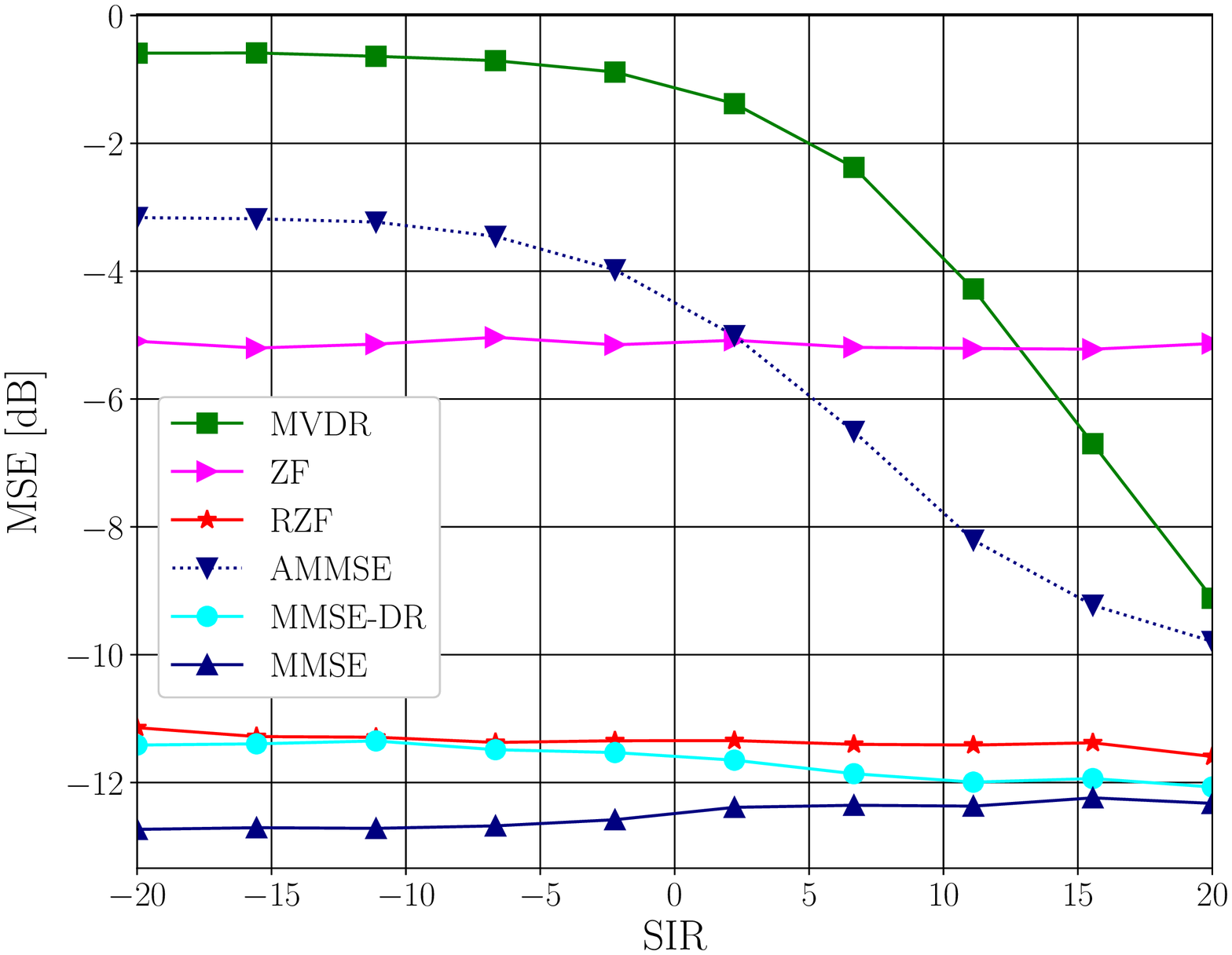}} &
\hspace*{-2em}
 \subfigure[$N=64$, $J=19$, SNR $=$ 0 dB]{
\includegraphics[width =0.52\textwidth]{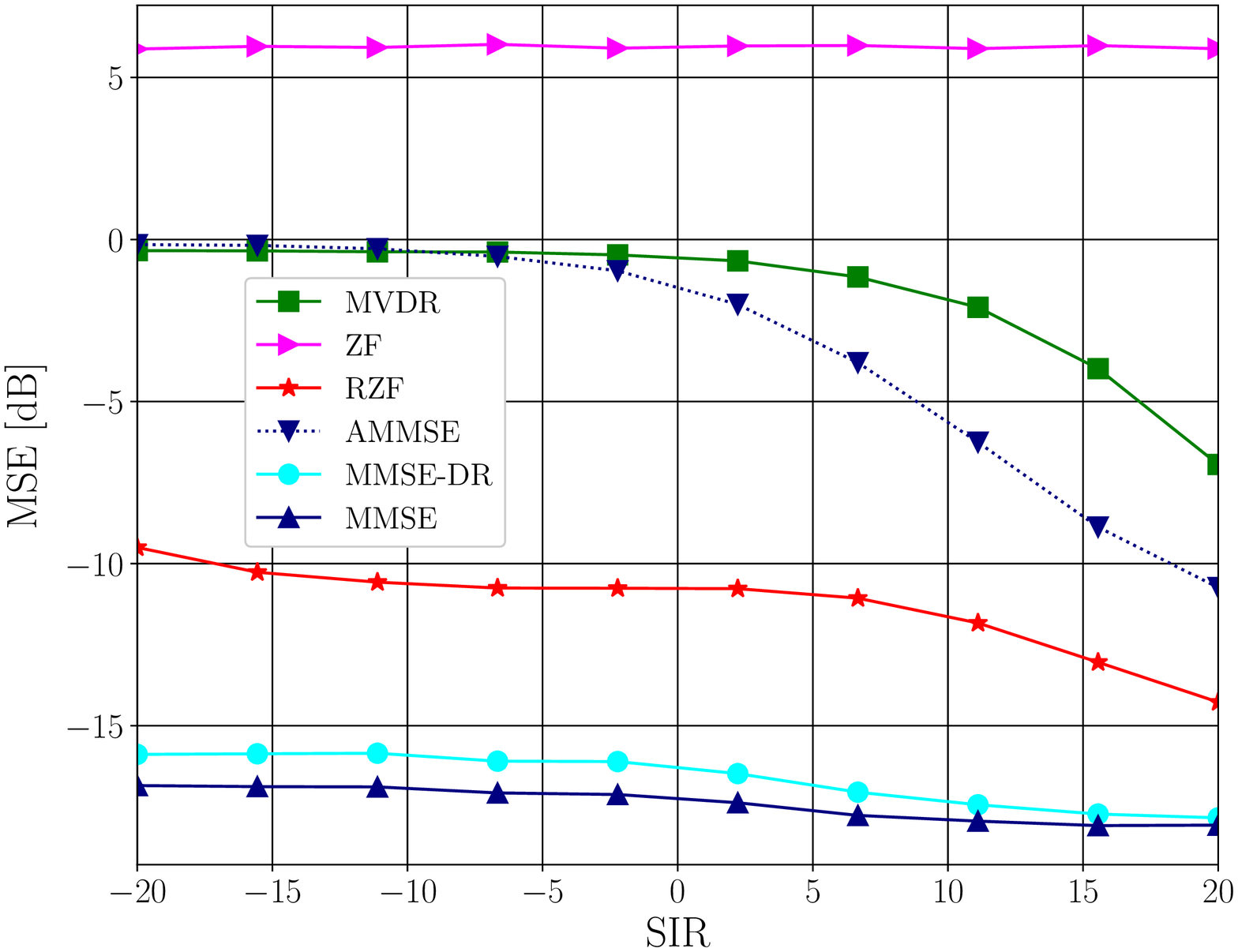}}
\end{tabular}

\begin{tabular}{cc}
\hspace*{-1.5em}
 \subfigure[$N=16$, $J=7$, SNR $=$ 10 dB]{
\includegraphics[width = 0.52\textwidth]{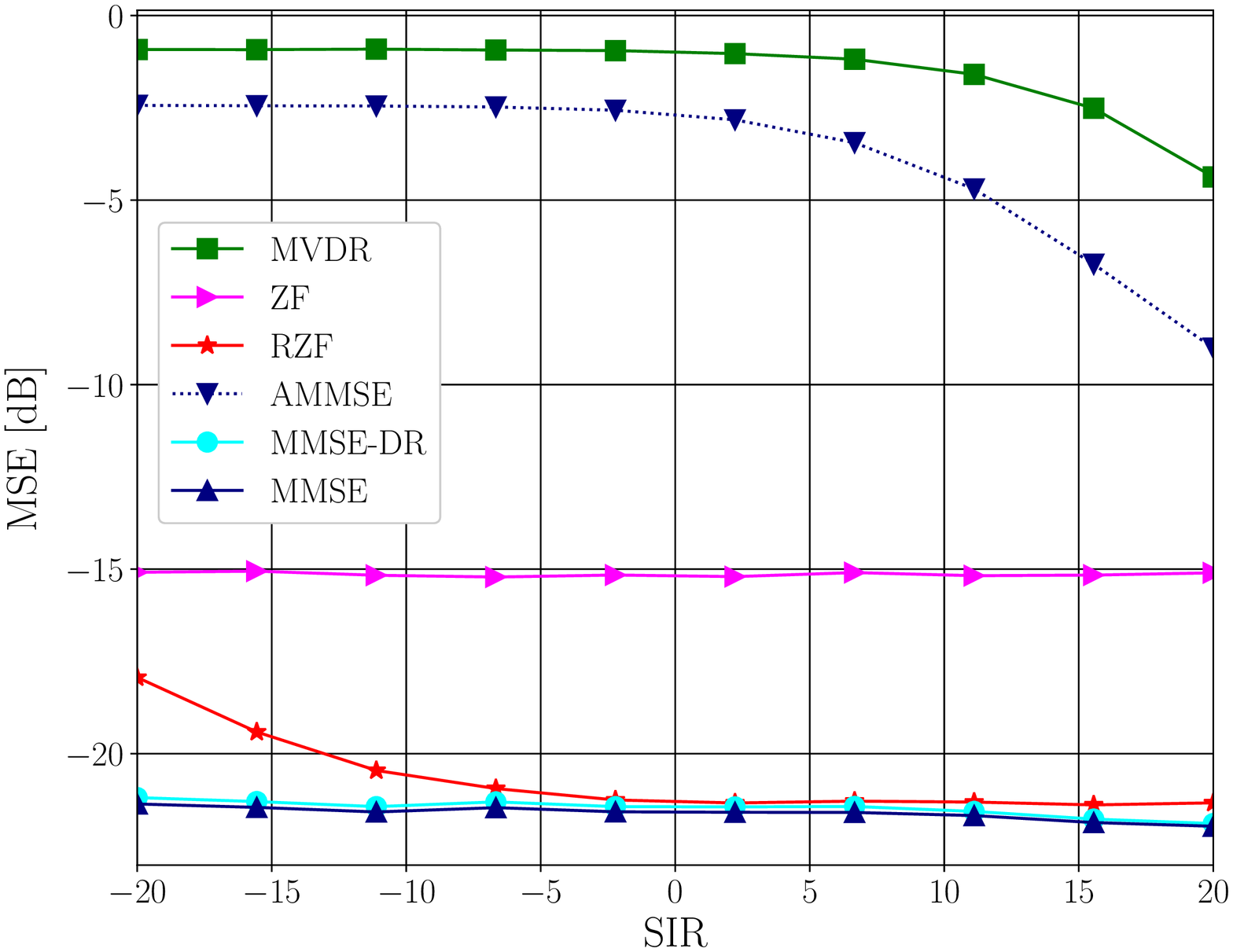}} &
\hspace*{-2em}
 \subfigure[$N=64$, $J=19$, SNR $=$ 10 dB]{
\includegraphics[width =0.52\textwidth]{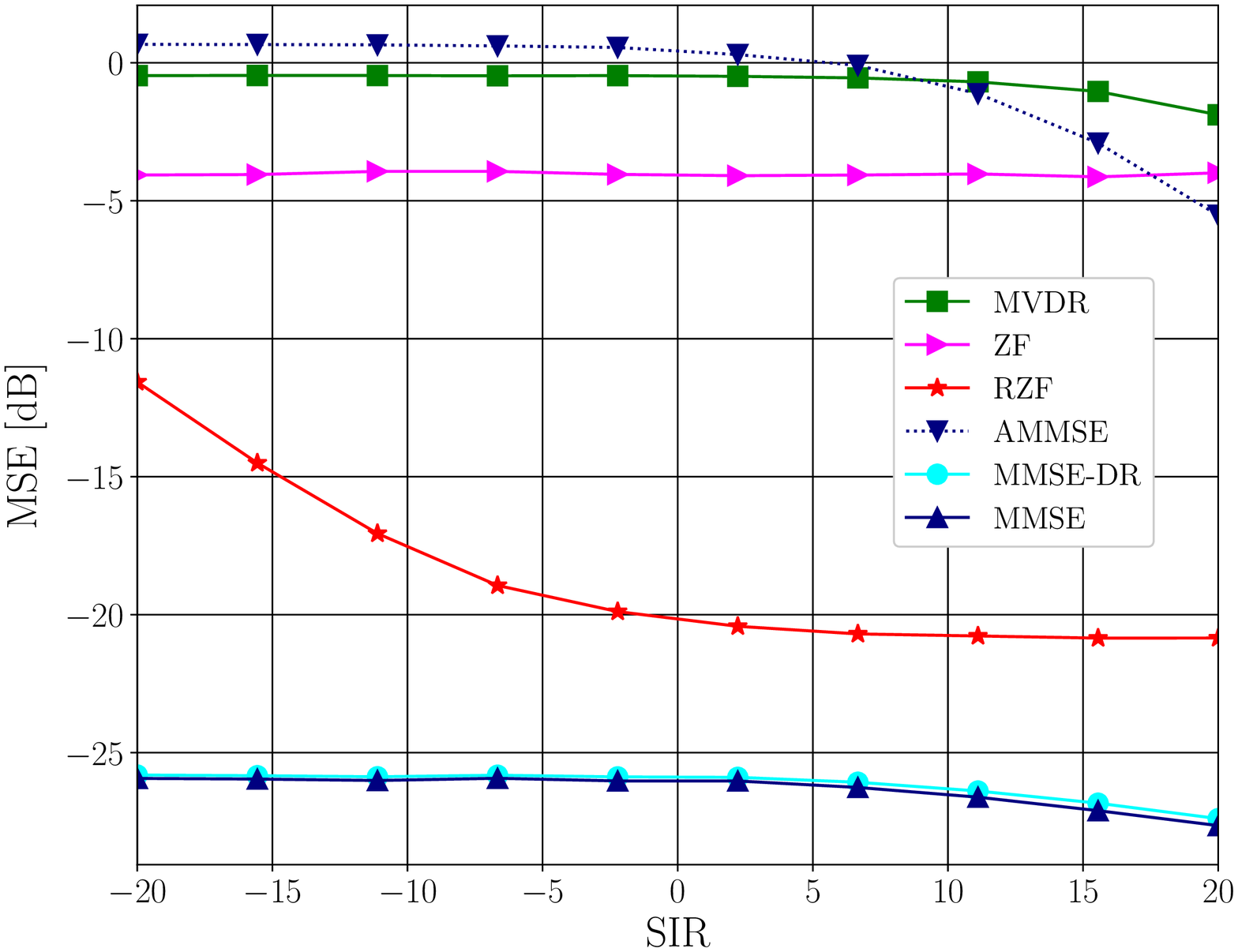}}
\end{tabular}

 \vspace{-5pt}
 	\caption{Performances  across SIR 
 for $\rho_j:=0.6$, $\beta:=0.8$,  $\varepsilon_\rho:=0.1$, $\varepsilon_\phi:=\pi/12$.}
 	\label{fig:different_sir}
 \end{figure}

A fresh look at the MSE expression in \eqref{eq:mse_cost_dr} may suggest that
one might ``exploit'' the temporal correlations to reduce the MSE further,
rather than bounding the leakage of interference.
To see whether this active approach works,
we consider an approximate MMSE (A-MMSE) beamformer under the assumption
that some erroneous estimates $\hat{c}_j\in \comp$ 
and $\hat{\sigma}_0^2>0$ of the correlation
$c_j(:=E[s_0^*(k)s_j(k)])$ and the signal power
$\sigma_0^2(:=E[\abs{s_0(k)}^2])$ are available.
To be precise,
the erroneous estimates $\hat{\sigma}_0^2$ and
$\hat{c}_j=\abs{\hat{c}_j}e^{i \hat{\phi}_j}$ are generated, respectively, by
$\hat{\sigma}_0^2:= \beta\sigma_0^2$ and
$\abs{\hat{c}_j} := \abs{c_j} + \varepsilon_\rho \sigma_0 \sigma_j$, and
$\hat{\phi}_j:=\phi_j + \varepsilon_\phi$,
where 
$\beta\in(0,+\infty)$,
$\varepsilon_\rho\in(-\rho_j,1-\rho_j)$, and
$\varepsilon_\phi\in(-\pi,\pi]$.
In this case, 
being free from the distortionless constraint,
the MSE $J_{{\rm MSE}}(\signal{w})$ in
 \eqref{eq:mse_cost} can be approximated by
$\hat{J}_{{\rm MSE}}(\signal{w}) 
:= \signal{w}^{\sf H} \signal{R} \signal{w}
+ \hat{\sigma}_0^2
(1-\signal{w}^{\sf H}\signal{h}_0-\signal{h}_0^{\sf H}\signal{w})
 - \sum_{j=1}^{J} (\hat{c}_j
 \signal{w}^{\sf H}\signal{h}_j + \hat{c}_j^*
 \signal{h}_j^{\sf H}\signal{w})$,
of which the minimizer is given by
$\signal{w}_{\rm A\mathchar`-MMSE}:= \signal{R}^{-1}
\left(\hat{\sigma}_0^2\signal{h}_0 + \sum_{j=1}^{J} \hat{c}_j \signal{h}_j\right)$.
The matrix $\signal{R}$ is computed from 8,000 samples for all the beamformers.

Figure \ref{fig:change_rhoj} depicts the performance of the beamformers
under different temporal-correlations for SNR 0 dB and SIR 0 dB.
It is seen that RZF exhibits remarkably better performances than MVDR 
in the strong correlation cases
(cf.~Remark \ref{remark:theorem_complex_mse}.5).
See Section \ref{subsec:discussions} for discussions about the gap
between the MSEs of RZF and MMSE-DR.
Figure \ref{fig:rzf} shows how the performance of RZF changes 
according to the choice of $\varepsilon$ for SNR 0 dB and SIR 0 dB.
It is seen that a wide range of the $\varepsilon$ parameter yields
lower MSEs than the conventional MVDR and ZF beamformers.
Figure \ref{fig:approximate_mmse} shows how the performance of A-MMSE 
changes due to the estimation errors in $\hat{c}_j$ and
$\hat{\sigma}_0^2$
for $N=16$, SNR 0 dB, and SIR 0 dB.
In Figure \ref{fig:approximate_mmse}(a),
$\varepsilon_\phi:=0$ and $\beta:=1$.
In Figure \ref{fig:approximate_mmse}(b),
$\varepsilon_\rho:=\varepsilon_\phi:=0$.
In contrast to the case of RZF, the performance of A-MMSE degrades by slight
errors in the estimates.
Figures \ref{fig:different_snr} and \ref{fig:different_sir} 
depict the performances for $\rho_j:=0.6$
under different SNR conditions with SIR 0 dB and 10 dB, and 
under different SIR conditions with SNR 0 dB and 10 dB, respectively.
The results show that RZF yields near optimal performances
in the case of $N:=16$ (excluding the low-SIR case in Figure \ref{fig:different_sir}(c)),
and 
that it outperforms MVDR and ZF significantly
over the whole settings in the case of $N:=64$.
Note here that the performance of RZF for $N:=64$ is suboptimal 
because the large $J$ makes the minimal angular separation between the desired
interfering sources be rather small.
Comparing the performances of RZF for
SIR $=$ 0 dB and 10 dB in Figure \ref{fig:different_snr},
the tendency is nearly the same; 
precisely, for high SNR,
closer performance to MMSE is achieved
in the case of SIR $=$ 10 dB 
than the case of SIR $=$ 0 dB.
Turning our attention to Figure \ref{fig:different_sir},
we observe that
(i) the maximum gain to MVDR is 10 dB approximately for SNR $=$ 0 dB
while it is 20 dB approximately for SNR $=$ 10 dB, and
(ii)  MSE of RZF increases towards that of ZF as SIR decreases,
more clearly for SNR $=$ 10 dB.
The first observation is due to the fact that RZF as well as ZF benefits
significantly from the reduction of noise.
The second one is also reasonable
since the zero-forcing strategy (i.e., $\epsilon:=0$) will be optimal
as the interference power approaches infinity.

\subsection{EEG Application}
\label{subsec:eeg}

Using the interpretation of the model in \refeq{eq:eeg_model} as 
the forward model used to solve the EEG inverse problem (\emph{cf.} Section~\ref{sec:prelim}), 
we consider the situation when
there are $J+1$ dipole sources $(J:=29)$ of brain activity and
the signals are measured by an array of $N:=128$ EEG sensors.
HydroCel Geodesic Sensor Net model is used as a realistic EEG cap layout. 
The source signals and forward model were generated using
SupFunSim library \cite{rykaczewski21}, which utilizes FieldTrip toolbox \cite{FieldTrip2011} for
generation of volume conduction model and leadfields.
The activity $s_0(k)$ of the desired source 
is generated by an autoregressive (AR) model of order $6$,
where the coefficients for each order are set to $0.2$.
The interfering activities are generated 
by \eqref{eq:sjk_generation} with $\phi_j:=0$.
The source activities are supposed to change in time, 
while their positions and orientations are assumed known and 
remain the same during the measurement period.
The following two cases are considered:
the low correlation case ($\rho_j := 0.5$) and 
the high correlation case ($\rho_j := 0.95$).
Throughout the experiments, the power of the desired signal is fixed
to $\sigma_0^2:=1$,
and those of the interference and noise are changed depending on
SNR and SIR, respectively.
The relaxation parameter $\varepsilon$ of RZF is optimized
for each SNR and SIR.

\begin{figure}[t!]
		\centering
\begin{tabular}{cc}
\hspace*{-1.5em}
 \subfigure[$\rho_j := 0.5$]{
\includegraphics[width = 0.52\textwidth]{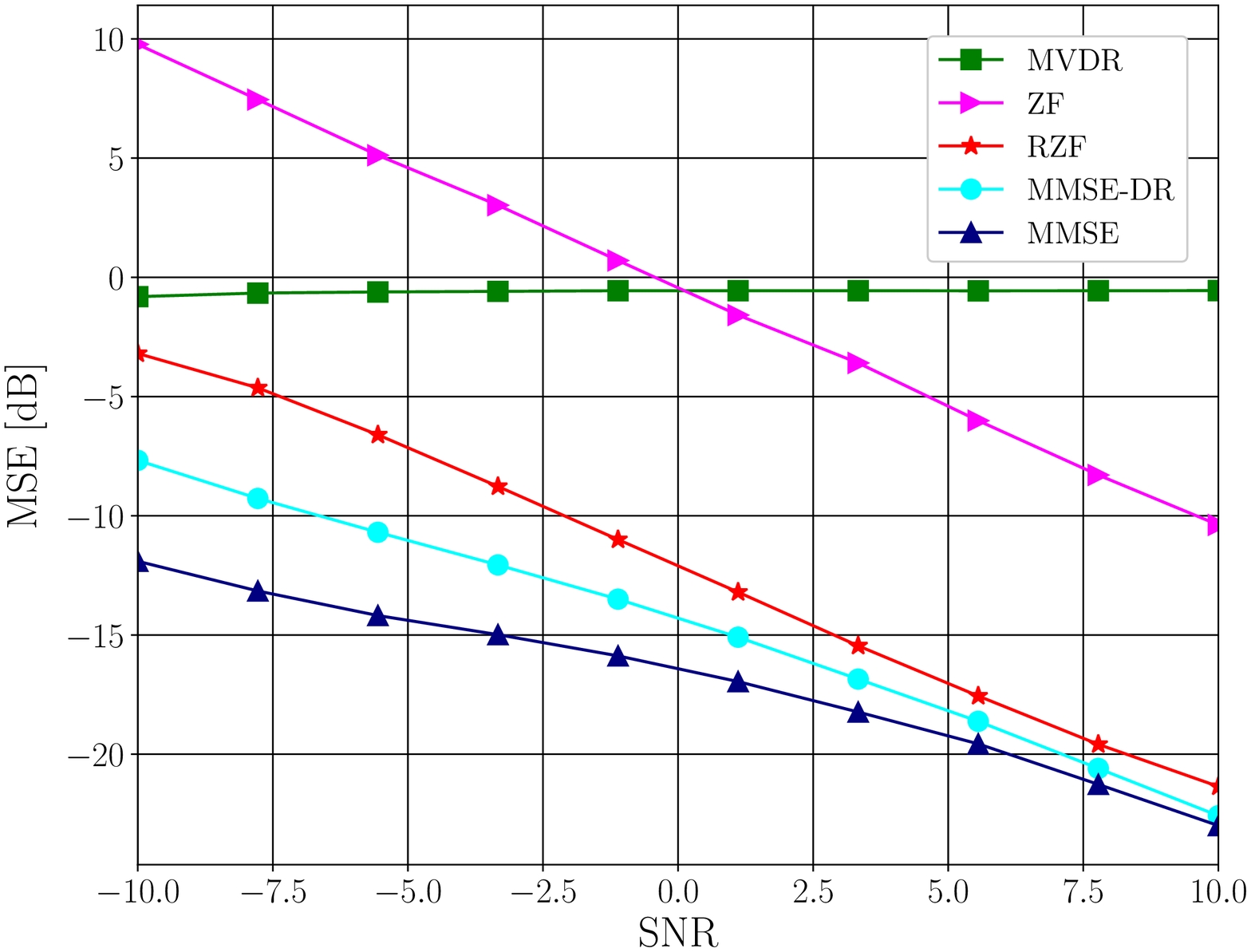}} &
\hspace*{-2em}
 \subfigure[$\rho_j := 0.95$]{
\includegraphics[width =0.52\textwidth]{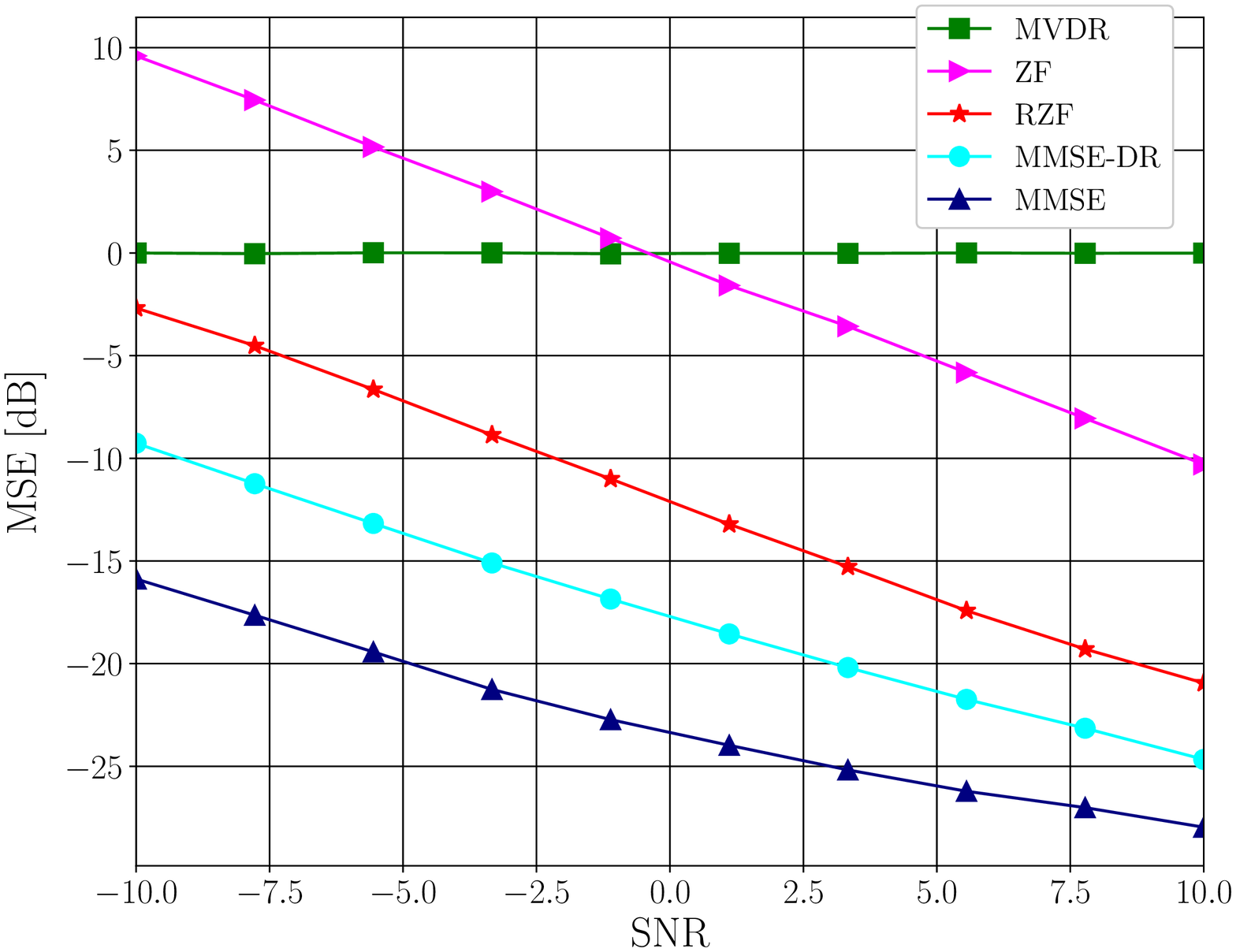}}
\end{tabular}

	\caption{Performances for EEG application under low and high correlations (SIR $0$ dB).}
	\label{fig: SNR_MSE}
\end{figure}

\begin{figure}[t!]
		\centering
		\centerline{\includegraphics[width = 0.52\textwidth]{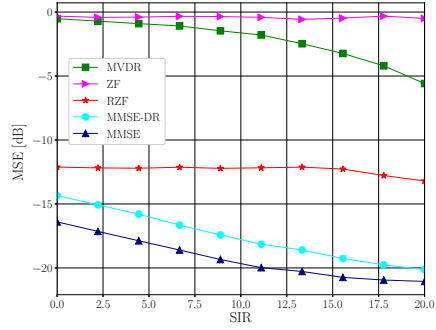}}
		\caption{Performance in EEG application under low correlation $\rho_j := 0.5$
	 (SNR $0$ dB).}	\label{fig: SIR_MSE}
\end{figure}

Figure \ref{fig: SNR_MSE} shows the results for SIR $0$ dB
under different SNR conditions
in the cases of (a) low correlation ($\rho_j:=0.5$) and 
(b) high correlation ($\rho_j:=0.95$).
Figure \ref{fig: SIR_MSE} shows the results for SNR $0$ dB
under different SIR conditions
in the case of low correlation ($\rho_j:=0.5$).
(The results for $\rho_j:=0.95$ are omitted as the results are similar.)
It is seen that the RZF beamformer attains significant gains compared to the
MVDR and ZF beamformers.
It is also seen that its performance is fairly close to 
the theoretical bound (MMSE-DR) in the low correlation case with low SIR.
See Section \ref{subsec:discussions} for discussions about the MSE gap
between RZF and MMSE-DR together with that for the wireless-network case.

\begin{figure}[t!]
		\centering
\subfigure[$\rho_j := 0.5$]{\includegraphics[width=.3\textwidth]{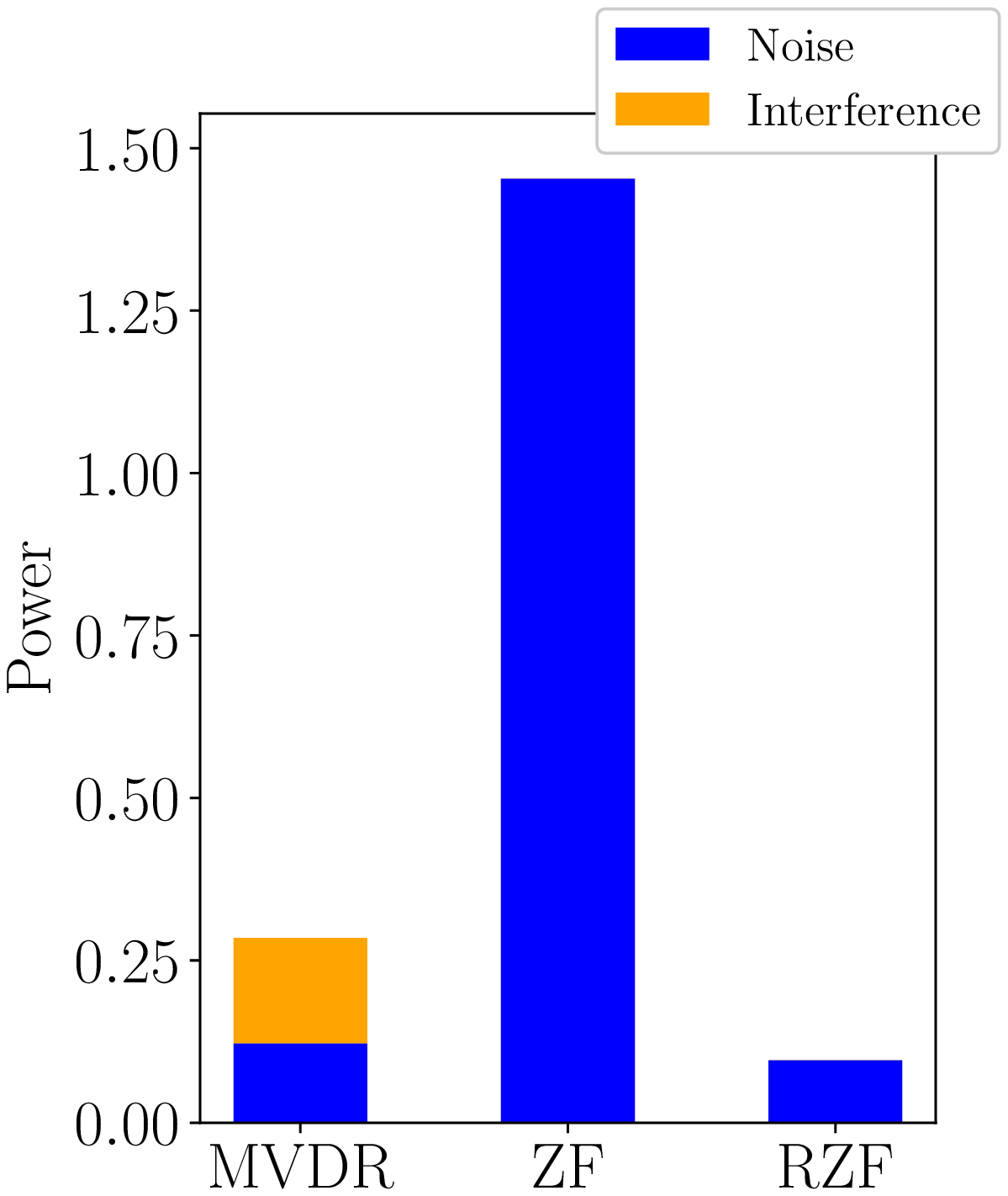}}
\subfigure[ $\rho_j := 0.95$]{\includegraphics[width=.3\textwidth]{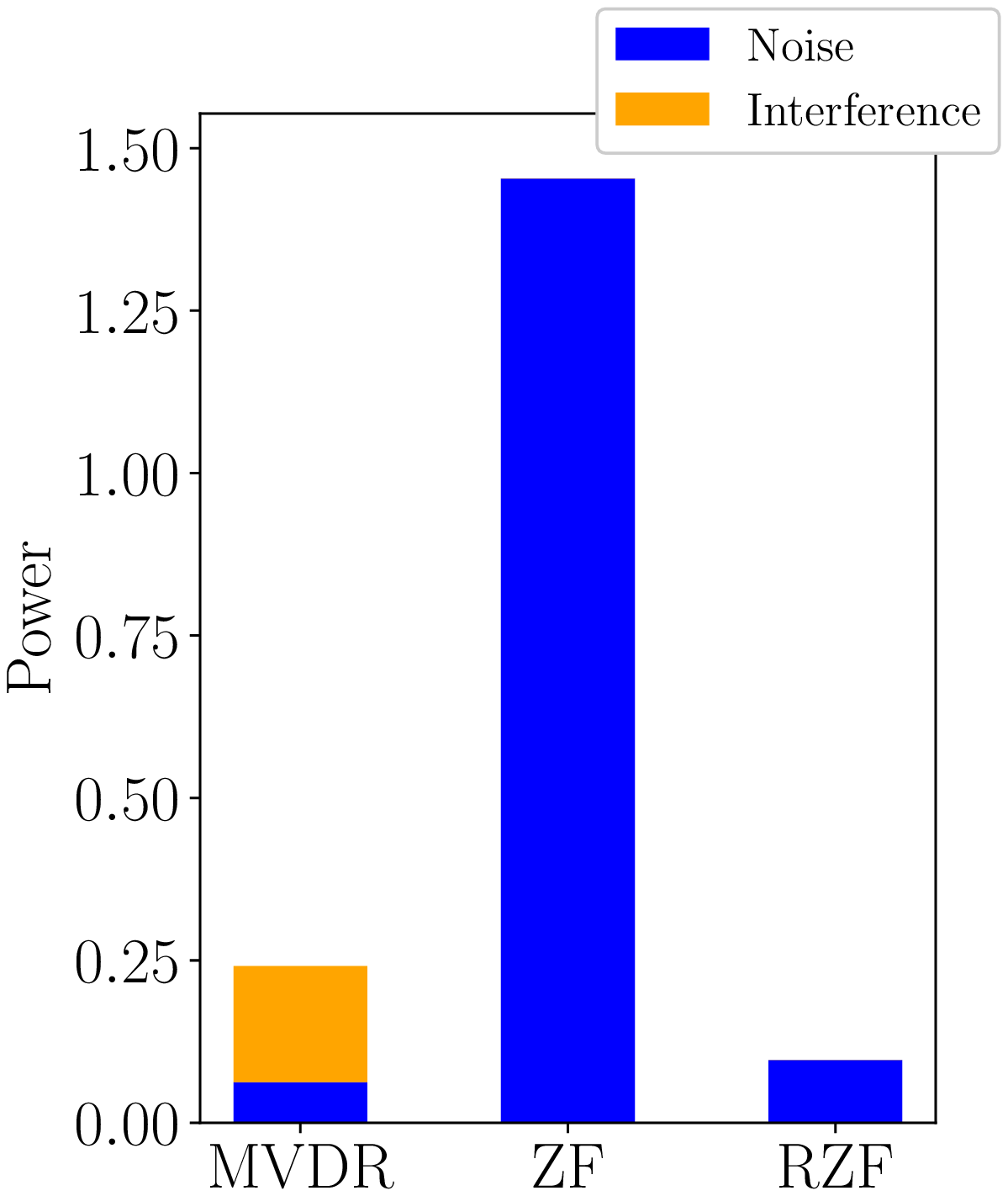}}
	\caption{Powers of noise/interference leakage for EEG application 
under low and high correlations (SNR $-2$ dB, SIR $0$ dB).}\medskip
	\label{fig: bar}
	\setlength{\belowcaptionskip}{-1em}
			\vspace{-0.5em}
\end{figure}

\begin{figure}[t!]
		\centering

\begin{tabular}{cc}
\hspace*{-1.5em}
 \subfigure[]{
 \psfrag{e}[Bl][Bl][0.9]{$\varepsilon$}
\includegraphics[width = 0.52\textwidth]{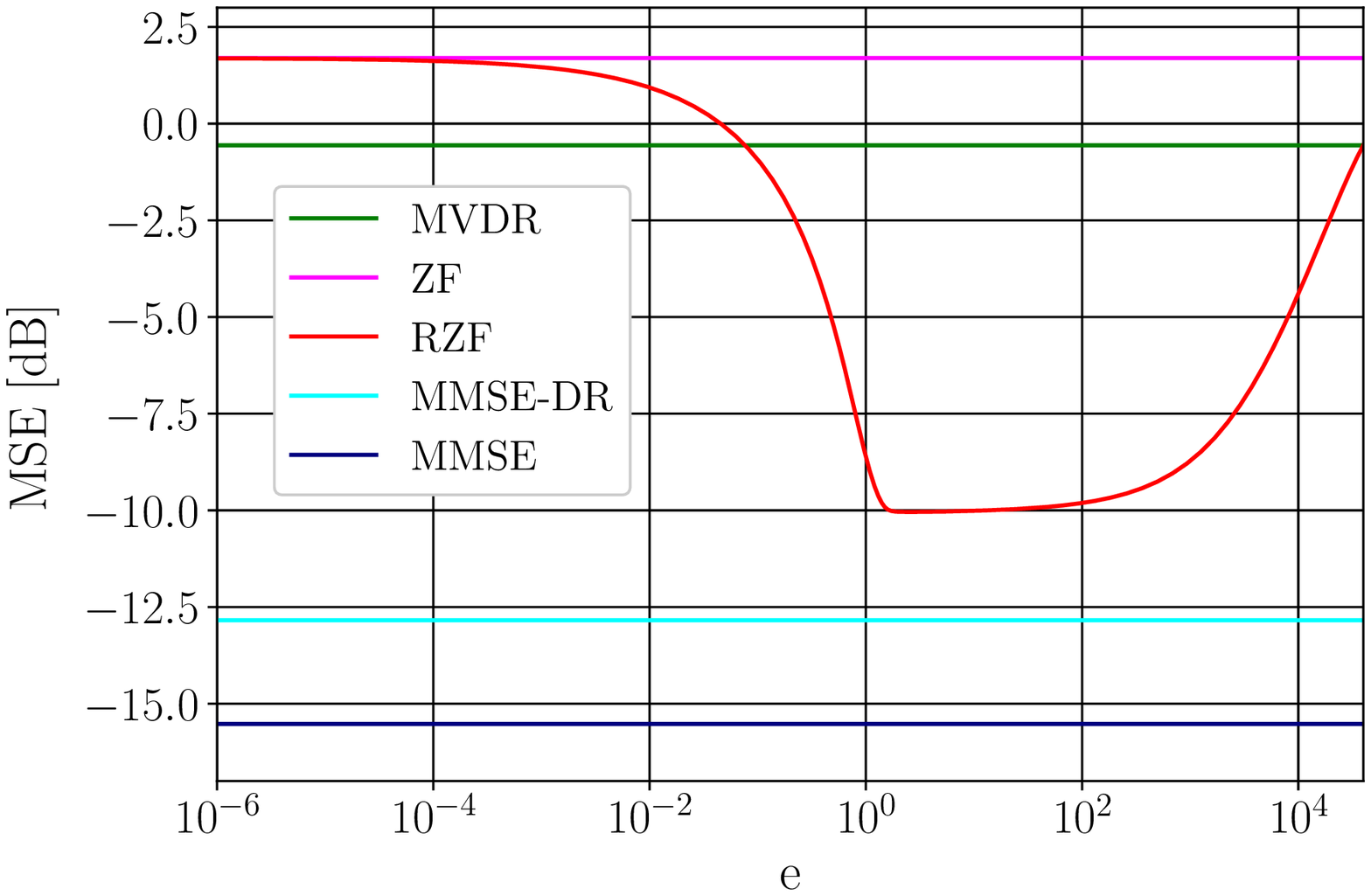}} &
\hspace*{-2em}
 \subfigure[]{
 \psfrag{a}[Bl][Bl][0.9]{$\varepsilon$}
\includegraphics[width =0.52\textwidth]{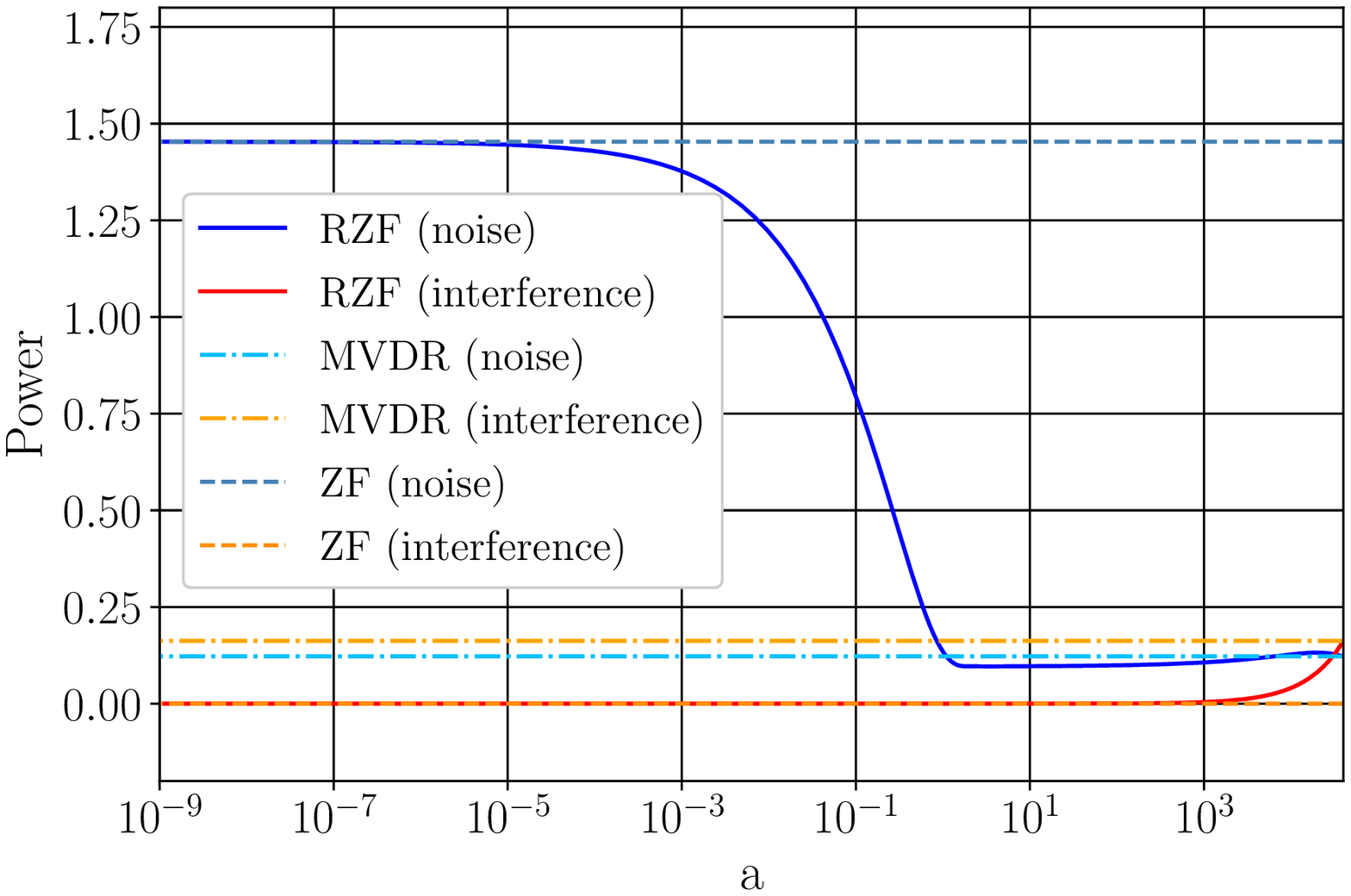}}
\end{tabular}

		\caption{Sensitivity of RZF to the parameter $\varepsilon$ 
for EEG application
($\rho_j := 0.5$, SNR $-2$ dB, SIR $0$ dB).}
		\label{fig: epsilon_MSE}
\end{figure}

Figure \ref{fig: bar} shows the powers of noise and interfering activities
remaining in the beamformer output
for SNR $ -2$ dB and SIR $0$ dB. 
It is seen that
RZF attains an excellent tradeoff;
it attains reasonably small noise-leakage
by allowing some slight (invisible) leakage of the interfering activities.
In Figure \ref{fig: bar}(a), 
the total leakage of RZF is slightly lower than the noise leakage of
MVDR (see Remark \ref{remark:noiseamp_real}).
Figure \ref{fig: epsilon_MSE}(a) plots the MSE performance of RZF
for each relaxation parameter $\varepsilon$
under SNR $-2$ dB, SIR $0$ dB, and $\rho_j:=0.5$.
Within the range of $ 1.0 <\varepsilon< 1.0 \times 10^{3}$,
the MSE of RZF is below $-7.5$ dB.
This implies that RZF is reasonably insensitive to the choice of $\varepsilon$.
Figure \ref{fig: epsilon_MSE}(b) provides more precise information,
plotting the power of noise/interference leakage contained in the beamformer output for different $\varepsilon$ values.
It is seen that both noise and interference are suppressed
simultaneously by RZF for an appropriately chosen $\varepsilon$.
This supports the results of Figure \ref{fig: bar}(a).

We now consider the adaptive implementations:
RZF implemented by DDAA, and MVDR and ZF implemented by 
CNLMS
\cite{apolinario1998constrained}.
The step sizes for all online algorithms are set to
$0.1$.
The parameter of the DDAA is set to
$\alpha_k:=0.5$, $\forall k\in\Natural$ (see the appendix).
Figure \ref{fig: Iter_MSE}(a) plots the steady-state MSE of each 
algorithm for SIR $0$ dB and $\rho_j:=0.5$ under different SNR conditions.
One can see that 
each beamformer is successfully implemented adaptively
by each adaptive algorithm.
Figure \ref{fig: Iter_MSE}(b) plots the learning curves
for the specific case of SNR $0$ dB, where each point is computed by
averaging the squared errors over 300 independent trials.
(For visual clarity, the MSE values are further averaged over the previous 30 iterations.)
It can be seen that the MSEs of the adaptive algorithms for the RZF and ZF
beamformers converge reasonably fast to
those of the analytical solutions, respectively.
In contrast, MVDR implemented by CNLMS converges slowly 
due to no use of the channel information of the interfering activities.

\begin{figure}[t!]
		\centering
\begin{tabular}{cc}
\hspace*{-1.5em}
 \subfigure[steady-state performance]{
\includegraphics[width = 0.52\textwidth]{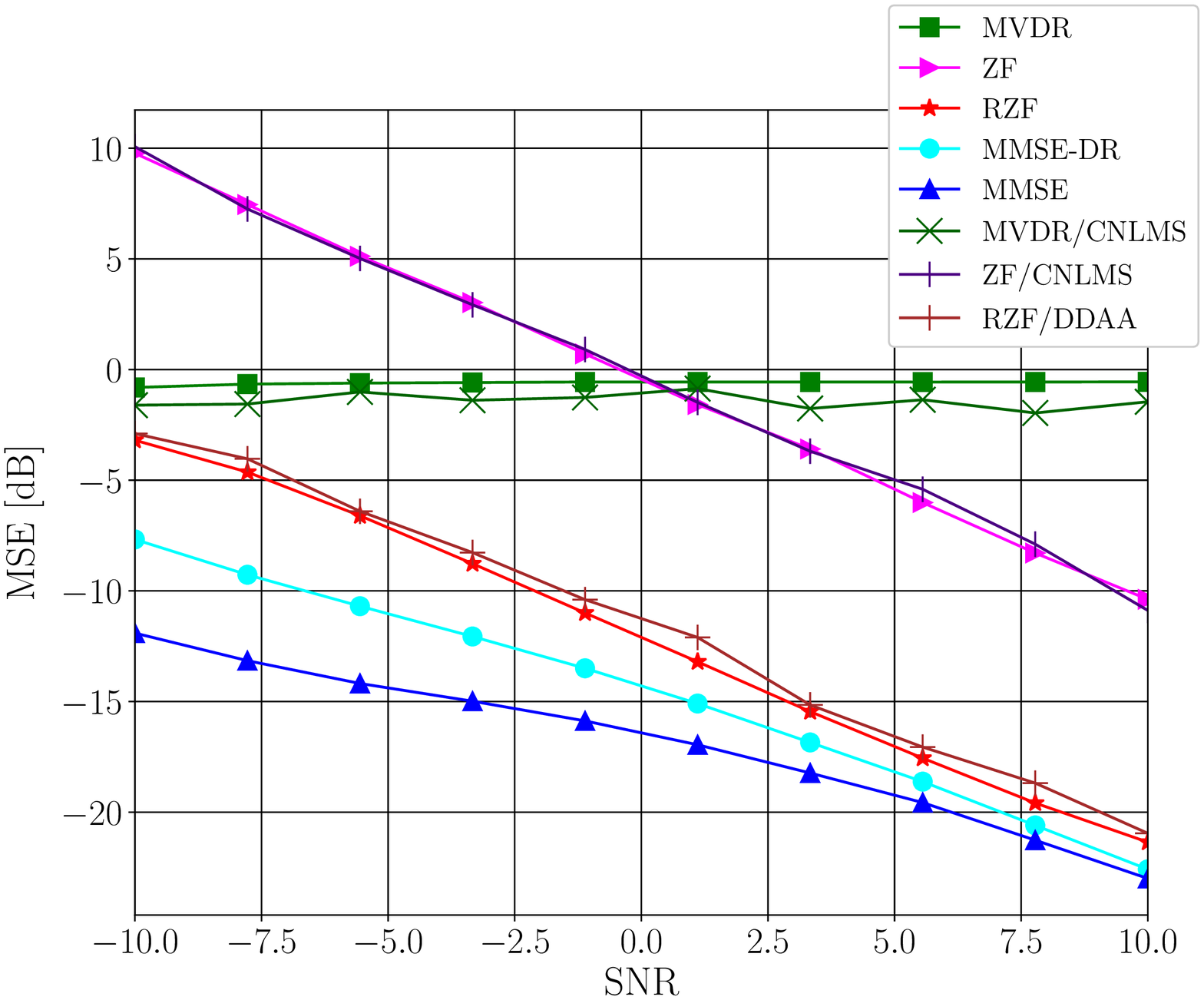}} &
\hspace*{-2em}
 \subfigure[learning curves under SNR $0$ dB]{
\includegraphics[width =0.52\textwidth]{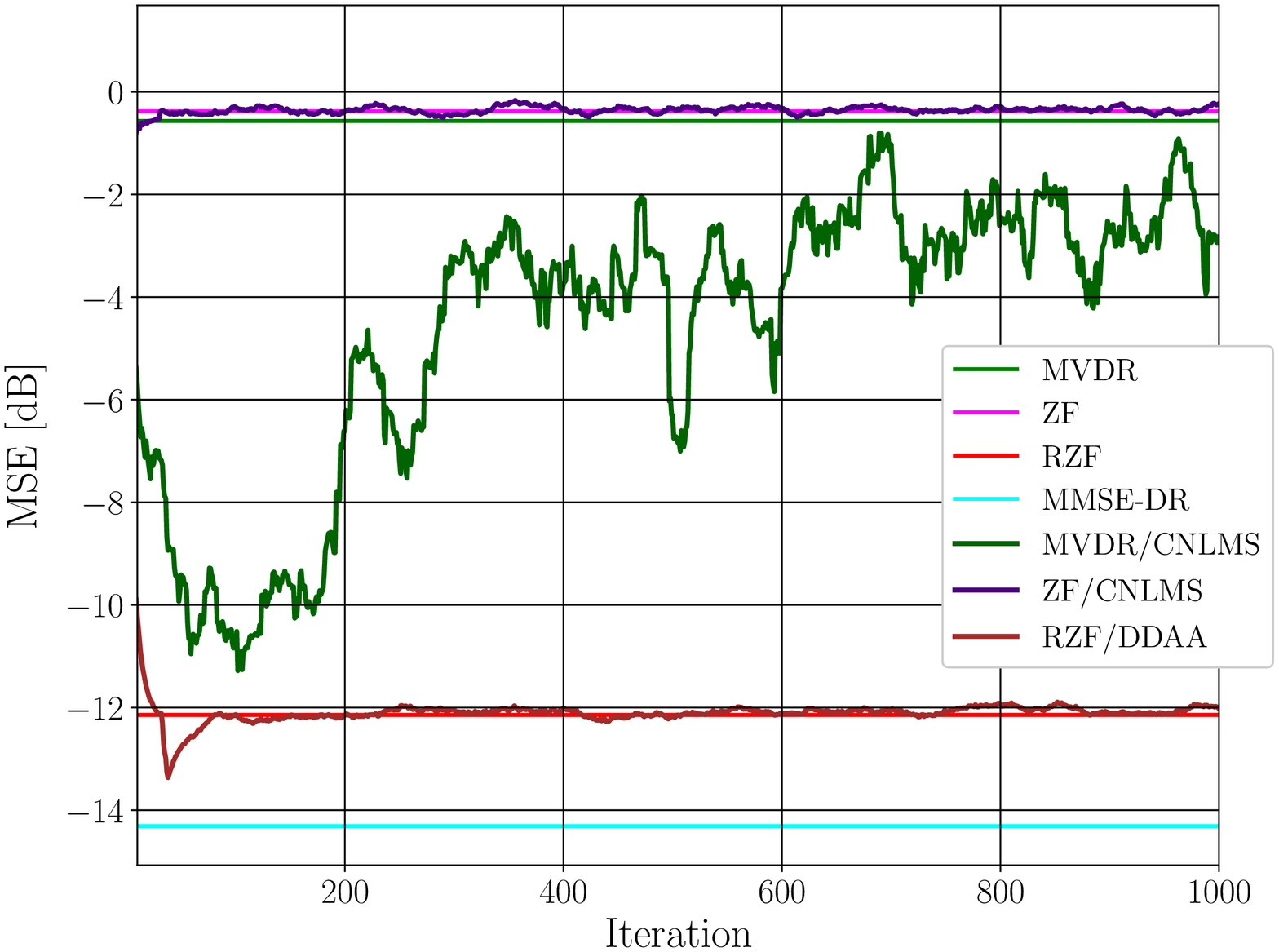}}
\end{tabular}

		\caption{Performances of adaptive implementations for EEG application
under low correlation $\rho_j := 0.5$ (SIR $0$ dB).
RZF implemented with DDAA, and
MVDR and ZF implemented both with CNLMS.}
		\label{fig: Iter_MSE}
\end{figure}

We remark finally that the RZF beamformer may be also used
when surface Laplacian-processed signals are considered instead of the
raw EEG signal to enhance spatial resolution \cite{murzin13,kayser15}.
In such a case, the channel matrix is replaced by the surface Laplacian
matrix
and the covariance matrix $\signal{R}$ is replaced by 
the covariance matrix of the Laplacian-treated signal.

\subsection{Discussions and Remarks}
\label{subsec:discussions}

\begin{figure}[t!]
\centering
\psfrag{g}[Bc][Bl][0.9]{$\gamma$}
\psfrag{i}[Bc][Bl][0.9]{Interfering source}
\centerline{\includegraphics[width =0.52\textwidth]{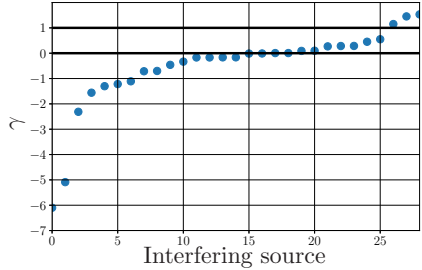}}
\caption{An example of the $\gamma$ value for
each interfering source in the EEG application 
under low correlation $\rho_j := 0.5$
(SNR $0$ dB, SIR $0$ dB).}
\label{fig:gamma_values}
\end{figure}

Figure \ref{fig:gamma_values} plots
the $\gamma$ values for different sources in the setting corresponding
to Figure \ref{fig: SIR_MSE} with SIR 0 dB.
It is seen that only 9 (out of 28) sources have the $\gamma$ values within
the range $(0,1)$; see \eqref{eq:mse_rzf_real} in Theorem \ref{theorem:real_mse}.
To see whether those interfering sources which have $\gamma$ 
outside the range $(0,1)$ cause
the MSE gaps between RZF and MMSE-DR,
we removed all such interfering sources
and evaluated the MSE performances
only with the remaining sources
having $\gamma\in(0,1)$
under $\rho_j:=0.5$, SNR $-5$ dB, and SIR $0$ dB.
The gap then droped,
from $3.4$ dB, down to $1.5$ dB approximately.
We also tested with the same number of {\it randomly picked}
sources removed.
In this case, the gap was no less than $2.8$ dB.
This indicates that 
those sources having $\gamma\not\in(0,1)$ would be
a main cause of the MSE gaps.
In the simulation results presented in Section \ref{subsec:communication},
there are some visible gaps between the MSEs of RZF and MMSE-DR
in Figure \ref{fig:change_rhoj}(b).
In the case of complex signals,
Theorem \ref{theorem:complex_mse} suggests
the existence of another factor
$\left(1-\abs{\delta_1/\delta_2}^2\right)
\sigma_n^4\tan^2\tau/(\sigma_1^2\cos^2\tau+\sigma_n^2)$
that may cause the MSE gap.

Indeed, $\gamma$ is a function of $\delta_1$, as well as the others.
Referring to \eqref{eq:delta1}, the sign of $\delta_1$ depends on the
phase $\phi_c$ of the {\it temporal} correlation $c_1$.
This implies that the MSE of RZF may change significantly
due to the phase (or the sign in particular) of $c_1$.
Nevertheless, the plots for RZF are nearly symmetric 
for the positive and negative values of $\rho_j(:=c_j/(\sigma_0\sigma_j))$.
In addition, we actually tested different values of $\phi_j$s, 
which yielded no visible differences in MSE.
This would be because the phase $\phi_{z_j}$ of the {\it spatial} correlation
$z_j:=\innerprod{\signal{h}_0}{\signal{h}_j}$ is quite random among different sources.

\section{Conclusion}

We studied the RZF beamformer, which minimizes the output variance
under the constraints of bounded interference leakage and undistorted
desired signal, in the presence of temporally correlated interference.
The (quadratic) {\it relaxed zero-forcing} constraint alleviates 
the gap between the output variance (available) and MSE (unavailable)
without amplifying the noise.
An adaptive implementation of 
the RZF beamformer based on the dual-domain adaptive algorithm
was also presented.
We analyzed the MSE of RZF for the single-interference case
and derived the formula of MSE
in terms of the spatio-temporal correlations
between the desired and interfering signals
as well as the variances of noise and interference,
clarifying that RZF achieves the minimum MSE
under the conditions.
The presented analysis gave 
useful insights into the multiple-interference case,
as discussed in the experimental section.
Numerical examples showed the remarkable advantages of RZF
over the MVDR and ZF beamformers.
The RZF beamformer also achieved near-optimal performance in some cases.
We conclude the present study by stating that the RZF beamformer will be useful
particularly for the EEG brain activity reconstruction
because
(i) it enjoys superior performance by mitigating the interfering sources efficiently,
(ii) it includes the MVDR and ZF beamformers as special cases, and 
(iii) it can be implemented efficiently by DDAA.

\appendix


\newcounter{appnum}
\setcounter{appnum}{1}

\setcounter{theorem}{0}
\renewcommand{\thetheorem}{\Alph{appnum}.\arabic{theorem}}

\setcounter{lemma}{0}
\renewcommand{\thelemma}{\Alph{appnum}.\arabic{lemma}}
\setcounter{example}{0}
\renewcommand{\theexample}{\Alph{appnum}.\arabic{example}}
\setcounter{equation}{0}
\renewcommand{\theequation}{\Alph{appnum}.\arabic{equation}}
\setcounter{claim}{0}
\renewcommand{\theclaim}{\Alph{appnum}.\arabic{claim}}
\setcounter{remark}{0}
\renewcommand{\theremark}{\Alph{appnum}.\arabic{remark}}

\section{Dual-Domain Adaptive Algorithm: An Implementation of RZF Beamformer}
\label{subsec:ddaa}

\vspace{7pt}
Given any closed convex subset $K$ of $\comp^m$ of
dimension $m$, the metric projection $P_K(\signal{x})$ of 
a point $\signal{x}\in\comp^m$ onto $K$ is defined by
$P_K(\signal{x}):=\argmin_{\signal{y}\in\comp^m}\norm{\signal{x}-\signal{y}}$.
The set $\{\signal{w}\in\comp^N \mid \|\signal{H}_{\rm I}^{\sf H}
\signal{w}\|^2 \leq \varepsilon\}$ is a (degenerate) ellipsoid
onto which the projection has no closed-form expression.
Fortunately, the projection of the dual-domain vector
$\signal{H}_{\rm I}^{\sf H} \signal{w}\in\comp^J$ onto the closed ball
$B_{\varepsilon}$ is easy to compute, thus used in the dual-domain algorithm.
Define the normalized channel matrix 
$\widetilde{\signal{H}}_{\rm I}:=\signal{H}_{\rm I}/\sigma_{1}(\signal{H}_{\rm I})$.
Given an initial beamforming vector $\signal{w}_{0} \in C$,
the DDAA update equation is then given as follows:
\begin{equation*}
\signal{w}_{k+1} := \signal{w}_{k} + \lambda_{\signal{w},k}\eta_{k}
\signal{g}_k,
\end{equation*}
where $\lambda_{\signal{w},k}\in (0,2)$ and
\begin{align}
\signal{g}_k:=&~
\alpha_{k}\signal{g}^{{\rm (1)}}_{k} + 
(1-\alpha_{k}) \signal{Q} \widetilde{\signal{H}}_{\rm I}\signal{g}^{{\rm (2)}}_k,
\nonumber\\
&~~~~~~~~\signal{Q}:=\signal{I}- 
\signal{h}_0\signal{h}_0^{\sf H}/(\signal{h}_0^{\sf H}\signal{h}_0),~~~
\alpha_{k}\in [0,1],
\nonumber\\
\signal{g}^{{\rm (1)}}_{k} :=&~ P_{C\cap H_k}(\signal{w}_{k}) - \signal{w}_{k}
= -\frac{\signal{w}_k^{\sf H} \signal{y}(k)}{\signal{y}(k)^{\sf H}
 \signal{Q} \signal{y}(k)}\signal{Q} \signal{y}(k),
\nonumber\\
&~~~~~~~~
H_k := \{\signal{w}\in \comp^{N} \mid
{\signal w}^{\sf H}\signal{y}(k) = 0\},
\nonumber
\end{align}
\begin{align}
\signal{g}^{{\rm (2)}}_{k} :=&~ P_{B_{\varepsilon}}\left( \widetilde{\signal{H}}_{\rm I}^{\sf H} \signal{w}_k\right) -
\widetilde{\signal{H}}_{\rm I}^{\sf H} \signal{w}_k
\nonumber\\
=&~ \begin{cases}
\Bigg(\dfrac{\sqrt{\varepsilon}}{\norm{\widetilde{\signal{H}}_{\rm I}^{\sf H} \signal{w}_k}} -1\Bigg)\widetilde{\signal{H}}_{\rm I}^{\sf H} \signal{w}_k,
 &
\mbox{if } \norm{\widetilde{\signal{H}}_{\rm I}^{\sf H} \signal{w}_k}^2
     > \varepsilon,
\\
\signal{0},& \mbox{if } \norm{\widetilde{\signal{H}}_{\rm I}^{\sf H}
     \signal{w}_k}^2 \leq \varepsilon,
    \end{cases}
\nonumber\\
\eta_k := &~
\begin{cases}
\dfrac{\alpha_{k}{\big \|}\signal{g}^{{\rm (1)}}_{k}{\big
\|}^{{\rm 2}} + (1-\alpha_{k}) \norm{\signal{g}_k^{(2)}}^2}
{\norm{\signal{g}_k}^2},
 & \mbox{if } \signal{g}_k\neq \signal{0},\\
1, & \mbox{otherwise.}
\end{cases}
\nonumber
\end{align}
Here, the vector $\signal{g}_{k}^{{\rm (1)}}$ contributes to reducing
the output variance in (\ref{eq:rzf1}),
while
$\signal{g}_{k}^{{\rm (2)}}$ contributes to reducing the violation of
the quadratic constraint of (\ref{eq:rzf2}).
The DDAA method has been proven to generate a sequence convergent to an asymptotically optimal
point under certain mild conditions. 
See \cite{ysy_ieice10_multi,yukawa2013dual} 
for the detailed properties of the algorithm.

\bibliography{adptl_rkhs,abf}

\end{document}